\documentclass[prd,twocolumn,superscriptaddress,nofootinbib]{revtex4-2}
\usepackage{epsfig}
\usepackage{amsmath}
\usepackage{amsfonts}
\usepackage{float}
\usepackage{amssymb}
\usepackage{slashed}
\usepackage{color}
\usepackage{pbox}
\usepackage{subfigure}
\usepackage[colorlinks,citecolor=black, linkcolor = black, urlcolor = black]{hyperref}
\usepackage{tabularx}
\usepackage{titlesec}
\usepackage{scrextend}
\usepackage[normalem]{ulem}

\def\lsim{\raise0.3ex\hbox{$\;<$\kern-0.75em\raise-1.1ex\hbox{$\sim\;$}}}
\def\gsim{\raise0.3ex\hbox{$\;>$\kern-0.75em\raise-1.1ex\hbox{$\sim\;$}}}

\newcommand{\be}{\begin{equation}}
\newcommand{\ee}{\end{equation}}
\newcommand{\bea}{\begin{eqnarray}}
\newcommand{\eea}{\end{eqnarray}}
\newcommand{\nn}{\nonumber}

\newcommand{\am}{\Delta a_\mu}

\graphicspath{{Figures/}}

\begin{document}

\title{Can leptophilic-ALP be a solution to the muon $(g-2)$ anomaly?}
\author{Sougata Ganguly}
\email{sganguly@cnu.ac.kr}
\affiliation{Department of Physics and Institute of Quantum Systems (IQS), 
Chungnam National University, Daejeon 34134, Republic of Korea}
\affiliation{School of Physical Sciences, Indian Association
for the Cultivation of Science, 2A $\&$ 2B Raja S.C. Mullick
Road, Kolkata 700032, India}
\author{Biswarup Mukhopadhyaya}
\email{biswarup@iiserkol.ac.in}
\affiliation{Department of Physical Sciences, Indian Institute of Science Education and Research,\\
Kolkata, Mohanpur - 741246, India}
\author{Sourov Roy}
\email{tpsr@iacs.res.in}
\affiliation{School of Physical Sciences, Indian Association
for the Cultivation of Science, 2A $\&$ 2B Raja S.C. Mullick
Road, Kolkata 700032, India}

\begin{abstract}
In the light of recent measurement of muon $(g-2)$, 
we investigate the phenomenological implications of an axion-like particle (ALP) which only couples to the standard model charged leptons.
We find that in a narrow mass range of ALP, it can alleviate the tension between the theoretical prediction and experimental observation of 
$(g-2)_\mu$ once we consider all possible one and two-loop diagrams. In particular, ALP can either explain the muon $(g-2)$ anomaly for $5\,{\rm GeV} \lesssim m_a \lesssim 6\,\rm GeV$
while satisfying the other experimental constraints, or be restricted by it.
\end{abstract}
\maketitle
\section{Introduction}
\label{sec:intro}
The precise measurement of the anomalous magnetic
moment of muon reveals the possibility of existing
physics beyond Standard Model (BSM). Combining
the old measurement of muon $(g-2)$
by BNL E821 experiment \cite{Muong-2:2006rrc} along with the recent one 
by Fermilab \cite{Muong-2:2023cdq}, it was observed that the experimental
observation of $a_\mu (\equiv (g-2)/2)$ differs with the Standard Model (SM)
prediction by more than $5\sigma$ and the deviation is given by \cite{Muong-2:2023cdq}
\bea
\am = (249 \pm 48)\times 10^{-11}\,\,\,.\nn
\eea

The apparent mismatch between theoretical prediction \cite{Czarnecki:2002nt,Melnikov:2003xd,Aoyama:2012wk,Gnendiger:2013pva,Kurz:2014wya,Colangelo:2014qya,
Davier:2017zfy,Masjuan:2017tvw,Colangelo:2017fiz,Keshavarzi:2018mgv,Colangelo:2018mtw,Hoferichter:2018kwz,Gerardin:2019vio,Bijnens:2019ghy,Colangelo:2019uex,Blum:2019ugy,Hoferichter:2019mqg,Aoyama:2019ryr,
Davier:2019can,Keshavarzi:2019abf, Czarnecki:2002nt,Melnikov:2003xd,Aoyama:2020ynm}
and experimental observation may be a pointer towards physics beyond the standard model (BSM).
People have proposed many new ideas to address this issue 
\cite{Baek:2001kca, Ma:2001md,Banerjee:2020zvi}.

It should, however be mentioned that the above discrepancy is based on the estimate of the hadronic
vacuum polarisation (HVP)/light-by-light (HLBL) contribution in data-driven approaches \cite{Aoyama:2020ynm}. On the other hand, estimates based on
lattice calculation exhibit closer statistical agreement \cite{Wittig:2023pcl}, although this may also in principle be due to theoretical
uncertainties in such estimates. It should also be mentioned 
that a stand-alone data-driven estimate, namely that by the CMD3 Collaboration \cite{CMD-3:2023alj}, 
has recently reported a rather satisfactory agreement with observations.

The discrepancy between the observation and theoretical prediction of $(g-2)_\mu$ is thus an 
issue which is still quite alive. Therefore, any theoretical scenario
beyond the standard model which has potential contribution to $(g-2)_\mu$, remains a topic of active interest. On the one hand, it may
be looked upon as a way of explaining $(g-2)_\mu$ if the discrepancy persists. On the other side, such a scenario gets subjected to serious constraints, 
if indeed the discrepancy turns out to be insignificant.

Another, \textit{prima facie} unrelated,  issue of the SM is the appearance of a CP violating phase ($\theta_{\rm QCD}$) in the quantum chromodynamics (QCD) Lagrangian 
\cite{PhysRevLett.37.8, PhysRevD.14.3432} which has to be
tightly constrained from the measurement of neutron electric dipole
moment and the upper bound of $\theta_{\rm QCD}$ is $10^{-10}$ \cite{PhysRevLett.97.131801}.
The problem of tiny value of $\theta_{\rm QCD}$ has been resolved
by proposing a new axial $U(1)$ global symmetry, known as 
Pecci-Quinn (PQ) symmetry \cite{Peccei:1977ur,Peccei:1977hh}.
In this scenario, $\theta_{\rm QCD}$ parameter is replaced by a dynamical field,
known as axion \cite{PhysRevLett.40.223, PhysRevLett.40.279}. 
The axion field generates tiny mass from the QCD non-perturbative effects
and it can act as a dark matter (DM) candidate \cite{Dine:1982ah,PRESKILL1983127,
ABBOTT1983133, MARSH20161,DILUZIO20201}. It is to be noted that the mass of axion and its 
coupling with SM are not free parameters because both of them are functions
of the PQ breaking scale. 

Generalising the idea of axion, one can also consider axion-like particles (ALP) 
which is a pseudo Nambu-Goldstone boson, appearing in the theory due to
the spontaneous breaking of an approximate global symmetry. An ALP can be motivated, for example,
in terms of some high-scale scenarios such as string theories \cite{Arvanitaki:2009fg, Cicoli:2012sz,Visinelli:2018utg}.
The phenomenology of ALP is rich since its mass and couplings
with the SM fields are independent parameters, in contrast to a PQ axion
\cite{Bauer:2017ris,Bauer:2019gfk,Calibbi:2020jvd,Bauer:2021mvw,Bertholet:2021hjl,Chakraborty:2021wda}. 
An ALP, according to some conjectures, may also serve as a DM \cite{Arias:2012az, Jaeckel:2014qea,Ho:2018qur} as well as a DM portal \cite{Hochberg:2018rjs}.
The couplings of ALP with the SM fields are similar to that of an axion
which implies that the ALP couples with two photon via a dimension five operator
and it can decay into two photon. 
The coupling of ALP with photon has very rich phenomenology and
it can be constrained from various laboratory, astrophysical observations as discussed in
\cite{Grossman:2002by,Cadamuro:2011fd,Jaeckel:2014qea,Bauer:2017ris,Chang:2018rso}.

In this work, we primarily focus on leptophilic ALP scenario where the ALP couples with the charged SM
leptons. The interaction of ALP with the SM is governed by the following
effective Lagrangian.
\bea
\mathcal{L}_a& =& 
\dfrac{1}{2}\partial^\mu a \partial_\mu a - \dfrac{1}{2}m_a^2 a^2
- \sum_{\ell} \dfrac{i c_{\ell \ell} m_\ell}{f_a} \bar{\ell}\gamma_5 \ell a\,,
\label{eq:lagrangian}
\eea
where the ALP $a$ of mass $m_a$ couples with the charged SM leptons $\ell$ of mass $m_\ell$ 
with coupling $m_\ell c_{\ell \ell}/f_a$ and $f_a$ is the ALP decay constant. In our analysis, 
we consider $c_{\ell\ell}/f_a$ coefficients are same for all the SM charged leptons.
Let us note that although the ALP does not couple with SM quarks and photon at tree level but these couplings
can be generated at loop level.

Considering Eq.\,\ref{eq:lagrangian} as our starting point, we consider all possible one and two loop
diagrams for the calculation of muon $(g-2)$. We found that in this set-up, muon $(g-2)$ anomaly can be explained
for $5\,{\rm GeV} \lesssim m_a \lesssim 6\,\rm GeV$ while satisfying the other constraints.
We also discuss the phenomenological consequences of
loop-induced ALP-photon and ALP-quark couplings. 

The rest of the article is organised as follows. In Section \ref{sec:g-2}, 
ALP interpretation of muon $(g-2)$ has been thoroughly discussed. The constraints
in the leptophilic-ALP scenario has been discussed in Section \ref{sec:constraints} and
finally we conclude in Section \ref{sec:conclu}.
\section{ALP interpretation of muon $(g-2)$ }
\label{sec:g-2}
\begin{figure}[h!]
\includegraphics[width = 0.48\textwidth]{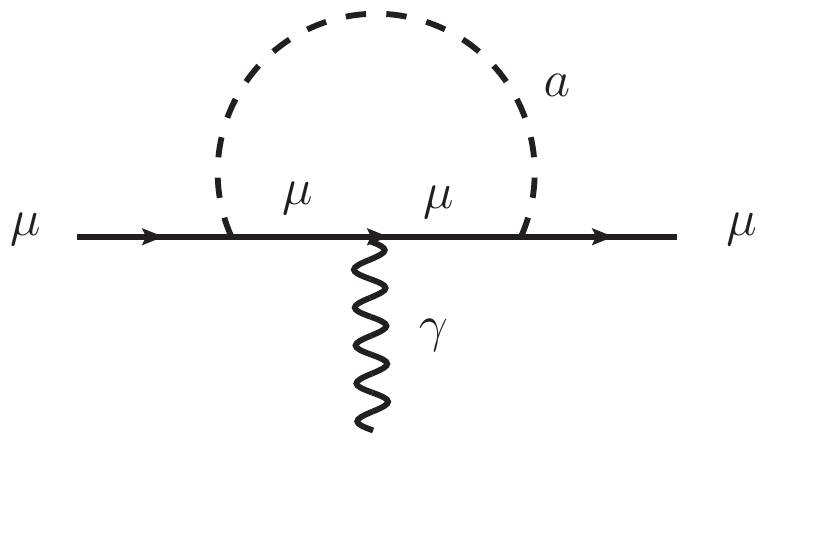}
\caption{One loop contributions to the anomalous magnetic moment of muon.}
\label{fig:1loop}
\end{figure}
The coupling of ALP with the muon opens up the possibility
to ameliorate the tension between SM prediction and experimental
observation of anomalous magnetic moment of muon.
The one loop contribution of ALP to the $\am$ (see Fig.\,\ref{fig:1loop})
is given by \cite{LEVEILLE197863,Lindner:2016bgg}
\bea
\am^{1-\rm loop}& =&  -\left(\dfrac{c_{\ell\ell} m_\mu}{f_a}\right)^2\dfrac{r}{8\pi^2}
\int_0^1 dx \dfrac{x^3}{1-x+rx^2}\,\,\,,
\label{eq:fig1}
\eea
where $m_\mu$ is the mass of the muon, $r = m_\mu^2/m_a^2$ and
the integral is always positive (since $r>0$ always).

As one can see from Eq.\,\ref{eq:fig1}, $\am^{1-\rm loop}$
is always negative irrespective of the
sign of $c_{\ell\ell}$ because of the presence of $\gamma_5$
in the ALP muon coupling. Thus, by considering Fig.\,\ref{fig:1loop} alone,
it is not possible to improve the theoretical prediction of $\am$.

\begin{figure}[hbt!]
\centering
\includegraphics[width = 0.4\textwidth]{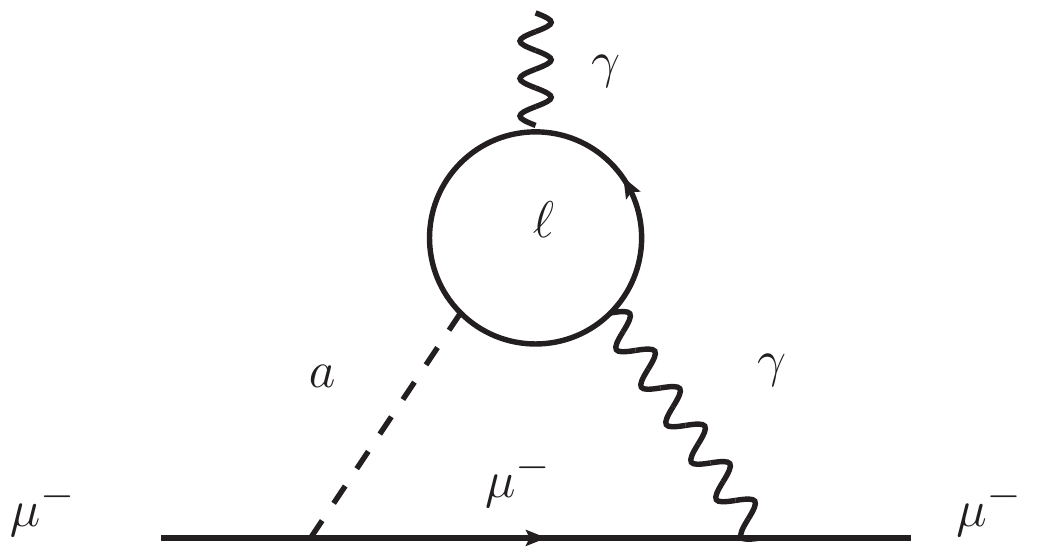}
\includegraphics[width = 0.4\textwidth]{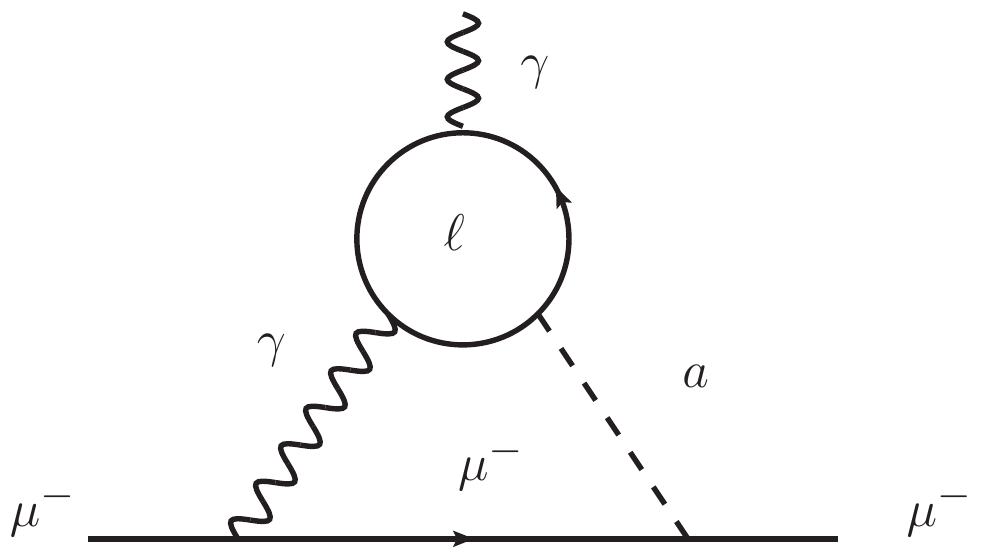}
\caption{Two loop contributions to the anomalous magnetic
moment of muon.}
\label{fig:2loop}
\end{figure}
However, in our framework, ALP-lepton couplings can contribute to the
muon $(g-2)$ at two-loop level.
The two loop diagrams are shown in Fig.\,\ref{fig:2loop} and its
contribution to $\am$ is given by the following expression \cite{Buttazzo:2020vfs}.
\bea
{\am}^{2-{\rm loop}} &=&
\dfrac{\alpha_{\rm em} m_\mu^2}{8\pi^3}
\left(\dfrac{c_{\ell \ell}}{f_a}\right)^2
\sum_\ell  q_\ell^2
\mathcal{F}\left(\dfrac{m_a^2}{m_\mu^2}, \dfrac{m_a^2} {m_\ell^2}\right)\,,\nn\\
\label{eq:fig2}
\eea
where $q_\ell$ is the electromagnetic charge
of the charged SM lepton $\ell$ running in the loop and the loop function
$\mathcal{F}(a, b)$ is given by
{\scriptsize
\bea
\mathcal{F} (a,b) = \int_0^1 dx dy dz
\dfrac{a x}{
a(1-x) + a b x y z (1-z) + b z (1-z)x^2 (1-y)^2}\,.\nn \\
\eea}
Let us note in passing, Eq.\,\ref{eq:fig2} always gives positive
contribution to $\am$ since $\mathcal{F} (a,b)>0,\,\,\forall \,a,\,b >0$.
It is to be noted, we have not considered the diagrams in which the photon
propagator is replaced by the $Z$ boson propagator. This is because, those
contributions are suppressed due to the massive $Z$ boson propagator and
$Z$ boson couplings with SM fermions.

Thus the total ALP contribution to $\am$ is given by
\bea
\am = \am^{1-\rm loop} + \am^{2-\rm loop}\,\,,
\eea
where individual expressions are given in Eq.\,\ref{eq:fig1}, and Eq.\,\ref{eq:fig2}. 

\begin{figure}[h!]
\includegraphics[width = 0.45\textwidth]{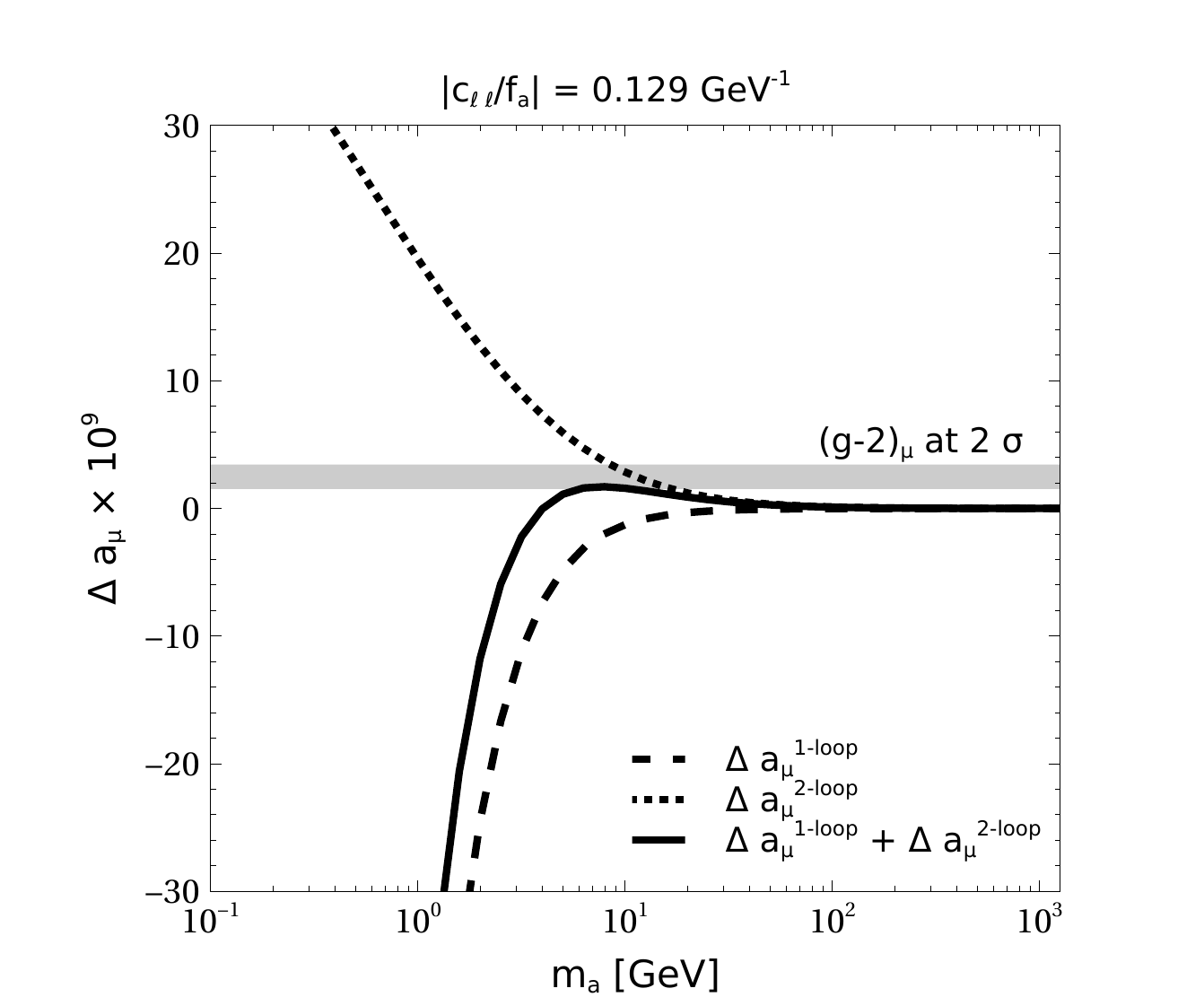}
\caption{Variation of $\am^{1-\rm loop}$ (dashed line), $\am^{1-\rm loop}$ (dotted line), and
$\am$ (solid line) as a function of ALP mass $m_a$ for $c_{\ell \ell}/f_a = 0.129 \, \rm GeV^{-1}$. The grey band denotes the 
$2\sigma$ limit of $\am$.}
\label{fig:BP1}
\end{figure}

In Fig.\,\ref{fig:BP1}, we show the effect of individual loop diagrams as a function of 
ALP mass for a fixed value of $c_{\ell \ell}/f_a$. As one can see, for 
$m_a \lesssim 1\,\rm GeV$, $\am < 0$ and 
$\am >0$ as we increase $m_a$ above a few GeV. This is because ALP contribution to $\am$ is a linear combination between a negative term, coming from Eq.\,\ref{eq:fig1}
and a positive term, coming from Eq.\,\ref{eq:fig2} and the 
interplay between these two contributions highly dependent on the ALP mass. 
One can clearly see from the figure that the negative contribution
gradually increases (decreases) as we decrease (increase)
$m_a$. Thus for small $m_a$ the positive contribution coming from Eq.\,\ref{eq:fig2} will be negated
by the large negative contribution. However, the effect of one loop contribution to 
$\am$ becomes smaller as we increase the ALP mass $m_a$ and as a result, 
positive contribution of Eq.\,\ref{eq:fig2} dominates over the negative
contribution of Eq.\,\ref{eq:fig1}. Finally, in Fig.\,\ref{fig:constraint}, we show the region of parameter 
space in $m_a - |c_{\ell \ell}/f_a|$ plane where the muon $(g-2)$ anomaly is satisfied within $2\sigma$ limit (red points).
In the next section, we will discuss the collider constraints in $m_a -c_{\ell \ell}/f_a$ plane.

\section{Existing constraints and result}
\label{sec:constraints}
In our proposed scenario, ALP-photon and ALP-quark couplings are absent at tree-level 
but these couplings can be generated radiatively, as mentioned in the introduction. 
Thus the constraints on the ALP-photon and ALP-quark coupling can be translated to 
the ALP-lepton coupling. The constraints on the ALP-lepton coupling from various 
observations are discussed below.

\begin{figure}
\includegraphics[width = 0.45\textwidth]{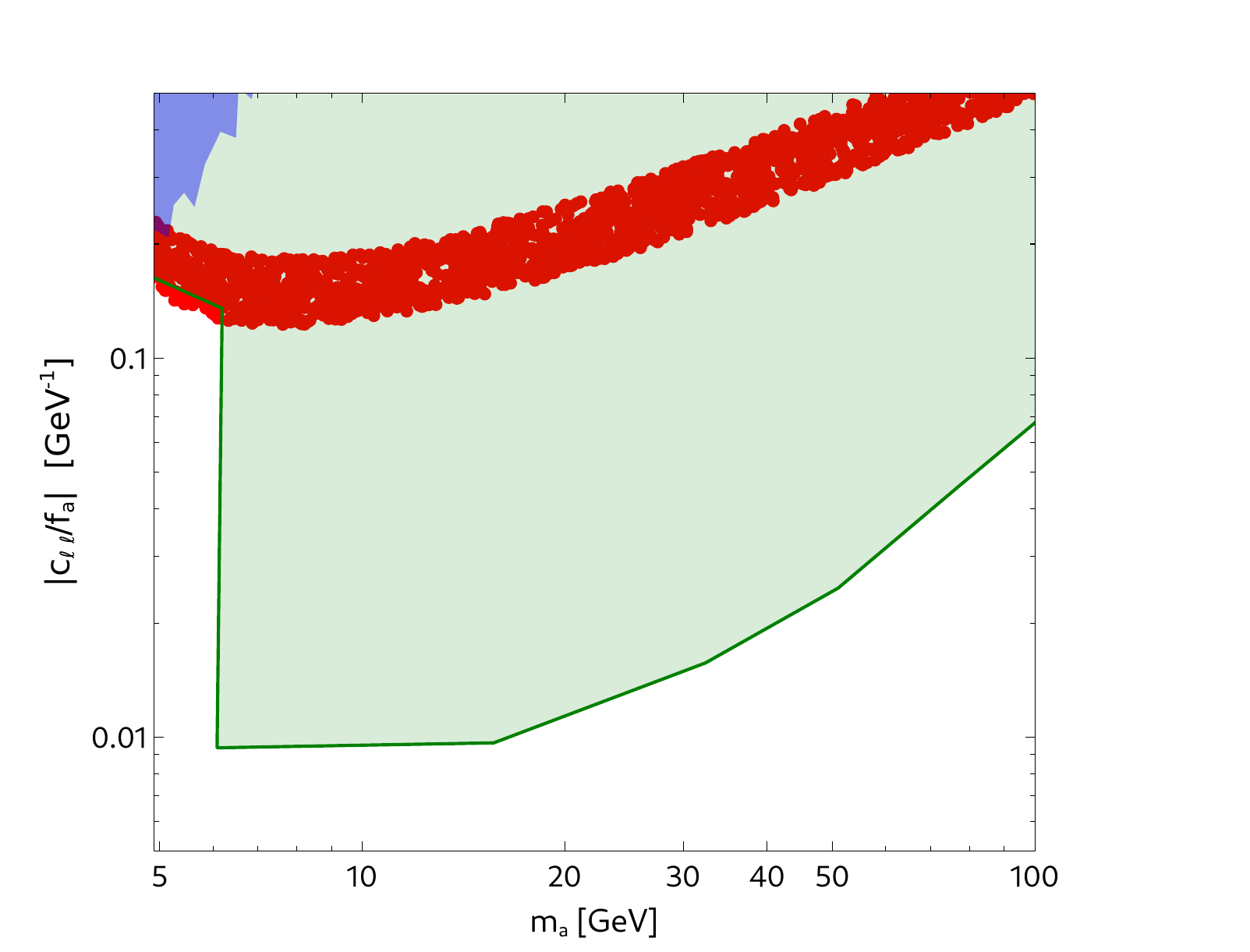}
\caption{ALP parameter space in $m_a$ - $|c_{\ell \ell}/f_a|$ plane. The red points are the $\am$ satisfied
points within $2\sigma$ limit. The light blue region is excluded from BaBar and the combined constraint from LEP, LHC, and CDF
is depicted by light green region.}
\label{fig:constraint}
\end{figure}

\begin{itemize}
\item Constraint from ALP-photon coupling: The ALP-photon coupling is severely constrained
from various astrophysical, cosmological, and laboratory observations \cite{Bauer:2017ris}. In our parameter
space of interest, the constraints mainly arise from the searches of 
$\gamma + \slashed{E}_T$ and tri-photon final states at LEP, LHC, and CDF \cite{Mimasu:2014nea, Jaeckel:2015jla}.
In our scenario, the ALP-photon coupling
can be generated at one-loop level and the effective ALP-photon coupling is given by \cite{Bauer:2017ris}
\bea
\dfrac{c_{\gamma \gamma}}{f_a} = \dfrac{c_{\ell \ell}}{16\pi^2 f_a}\sum_{\ell} {\cal B}(\tau_\ell)\,
\label{eq:cggbyfa}
\eea
where
$\tau_\ell = 4 m^2_\ell/m_a^2$ and the loop function ${\cal B} (\tau_\ell)$ is given by
\bea
{\cal B} (\tau_\ell) = 1 - \tau_\ell f(\tau_\ell)^2
\eea
with
\bea
f(\tau_\ell) &=& \arcsin \left(\dfrac{1}{\sqrt{\tau_\ell}}\right) \text{ for } \tau_\ell \ge 1 \nn\\
& =& \dfrac{\pi}{2} + \dfrac{i}{2} 
\ln \left(\dfrac{1 + \sqrt{1-\tau_\ell}}{1 - \sqrt{1-\tau_\ell}}\right)
\text{ for } \tau_\ell < 1\,\,.
\eea
Using Eq.\,\ref{eq:cggbyfa}, we translate the constraints from $m_a - c_{\gamma \gamma}/f_a$ plane
to $m_a -c_{\ell \ell}/f_a$ plane and these constraints are shown by the light green
region in Fig.\,\ref{fig:constraint}.
\item Constraint from ALP-lepton coupling: In our scenario, the tree-level ALP-lepton
coupling can be constrained from $e^+e^- \to \mu^+\mu^-a \to 2\mu^+2\mu^-$ search at BaBar \cite{BaBar:2016sci} and
the relevant constraint is shown by the light blue region of Fig.\,\ref{fig:constraint}. The constraint on the ALP parameter space from
the recently released data of $e^+e^- \to \mu^+\mu^-a \to 2\mu^+2\mu^-$ search at Belle II \cite{Belle-II:2024wtd} requires a dedicated analysis
which is beyond the scope of this work.

\item Constraint from ALP-quark coupling: Existence of ALP-quark coupling can induce
flavor-violating decay modes and thus the ALP-quark
coupling can be constrained from the upper limit of these decay modes , as discussed in 
\cite{Dolan:2014ska, Bauer:2017ris, Bauer:2021mvw}.

In our model, the ALP-quark coupling is absent at tree-level but it can be generated at two-loop
level \cite{Ametller:1983ec}. Parametrizing the ALP-quark coupling as $-i g_Y (\sqrt{2}m_q/v) \bar{q} \gamma_5 q$, the estimated
value of the ALP-top quark coupling for $m_a = 5\,\rm GeV$ is $g_Y \simeq 10^{-2} \times c_{\ell \ell}/f_a \rm GeV$.
Thus for $c_{\ell \ell}/f_a \sim 0.1 \rm GeV^{-1}$, the value of $g_Y$ is two-orders of magnitude smaller
than the existing constraint from $B_s \to \mu^+ \mu^-$.
\end{itemize}

\section{Summary and Conclusions}
\label{sec:conclu}
In this work we have studied the propsect of ALP to satisfy the muon $(g-2)$ anomaly.
In order to investigate this, we consider leptophilic-ALP model where the ALP couplings
with quarks and photon are absent at tree-level. In this formulation, we consider all possibile one
and two-loop diagrams for the calculation of muon $(g-2)$ anomaly. We found that in our model
it is possible to explain the muon $(g-2)$ anomaly due to presence of two-loop diagrams
because of its positive contribution to $\am$. We also discuss the phenomenological
implications of loop-induced ALP-photon and ALP-quark couplings. Due to these loop induced
effects, the constraints on the ALP-photon and ALP-quark couplings can be translated into
the ALP-lepton interaction. We found that for leptophilic-ALP scenario, most of the parameter space
is excluded by the existing collider bounds. However, in a narrow region of parameter space
between $m_a \simeq 5\,\rm GeV$ to $m_a \simeq 6\,\rm GeV$ muon $(g-2)$ anomaly can be explained by ALP
while satisfying the other existing constraints. Since this picture regarding the 
``estimate-vs-observation" issue on $(g-2)_\mu$ is still somewhat unclear, it is fair 
to say that the above statements apply to the situation where this shortfall in 
theoretical prediction persists. On the other hand, if the deficit finally 
disappears, and suggested by lattice estimates on the CMD3 analysis, then 
the game gets turned around. In such an eventuality, one might expect the 
`favoured' and `disfavoured' regions in the ALP parameter space to be 
swapped, subject to all other phenomenological constraints.

\section{Acknowledgements}
The work of SG is supported by the University Grants Commission (UGC), Government of India,
Indian Association for the Cultivation of Science (IACS), and the National Research Foundation of Korea (NRF-2022R1C1C1011840, NRF-2022R1A4A5030362).
We would like to thank Avirup Ghosh and Satyanarayan Mukhopadhyay for helpful comments.

\bibliography{references}

\begin{thebibliography}{65}%
\makeatletter
\providecommand \@ifxundefined [1]{%
 \@ifx{#1\undefined}
}%
\providecommand \@ifnum [1]{%
 \ifnum #1\expandafter \@firstoftwo
 \else \expandafter \@secondoftwo
 \fi
}%
\providecommand \@ifx [1]{%
 \ifx #1\expandafter \@firstoftwo
 \else \expandafter \@secondoftwo
 \fi
}%
\providecommand \natexlab [1]{#1}%
\providecommand \enquote  [1]{``#1''}%
\providecommand \bibnamefont  [1]{#1}%
\providecommand \bibfnamefont [1]{#1}%
\providecommand \citenamefont [1]{#1}%
\providecommand \href@noop [0]{\@secondoftwo}%
\providecommand \href [0]{\begingroup \@sanitize@url \@href}%
\providecommand \@href[1]{\@@startlink{#1}\@@href}%
\providecommand \@@href[1]{\endgroup#1\@@endlink}%
\providecommand \@sanitize@url [0]{\catcode `\\12\catcode `\$12\catcode
  `\&12\catcode `\#12\catcode `\^12\catcode `\_12\catcode `\%12\relax}%
\providecommand \@@startlink[1]{}%
\providecommand \@@endlink[0]{}%
\providecommand \url  [0]{\begingroup\@sanitize@url \@url }%
\providecommand \@url [1]{\endgroup\@href {#1}{\urlprefix }}%
\providecommand \urlprefix  [0]{URL }%
\providecommand \Eprint [0]{\href }%
\providecommand \doibase [0]{https://doi.org/}%
\providecommand \selectlanguage [0]{\@gobble}%
\providecommand \bibinfo  [0]{\@secondoftwo}%
\providecommand \bibfield  [0]{\@secondoftwo}%
\providecommand \translation [1]{[#1]}%
\providecommand \BibitemOpen [0]{}%
\providecommand \bibitemStop [0]{}%
\providecommand \bibitemNoStop [0]{.\EOS\space}%
\providecommand \EOS [0]{\spacefactor3000\relax}%
\providecommand \BibitemShut  [1]{\csname bibitem#1\endcsname}%
\let\auto@bib@innerbib\@empty
\bibitem [{\citenamefont {Bennett}\ \emph {et~al.}(2006)\citenamefont {Bennett}
  \emph {et~al.}}]{Muong-2:2006rrc}%
  \BibitemOpen
  \bibfield  {author} {\bibinfo {author} {\bibfnamefont {G.~W.}\ \bibnamefont
  {Bennett}} \emph {et~al.} (\bibinfo {collaboration} {Muon g-2}),\ }\bibfield
  {title} {\bibinfo {title} {{Final Report of the Muon E821 Anomalous Magnetic
  Moment Measurement at BNL}},\ }\href
  {https://doi.org/10.1103/PhysRevD.73.072003} {\bibfield  {journal} {\bibinfo
  {journal} {Phys. Rev. D}\ }\textbf {\bibinfo {volume} {73}},\ \bibinfo
  {pages} {072003} (\bibinfo {year} {2006})},\ \Eprint
  {https://arxiv.org/abs/hep-ex/0602035} {arXiv:hep-ex/0602035} \BibitemShut
  {NoStop}%
\bibitem [{\citenamefont {Aguillard}\ \emph {et~al.}(2023)\citenamefont
  {Aguillard} \emph {et~al.}}]{Muong-2:2023cdq}%
  \BibitemOpen
  \bibfield  {author} {\bibinfo {author} {\bibfnamefont {D.~P.}\ \bibnamefont
  {Aguillard}} \emph {et~al.} (\bibinfo {collaboration} {Muon g-2}),\
  }\bibfield  {title} {\bibinfo {title} {{Measurement of the Positive Muon
  Anomalous Magnetic Moment to 0.20~ppm}},\ }\href
  {https://doi.org/10.1103/PhysRevLett.131.161802} {\bibfield  {journal}
  {\bibinfo  {journal} {Phys. Rev. Lett.}\ }\textbf {\bibinfo {volume} {131}},\
  \bibinfo {pages} {161802} (\bibinfo {year} {2023})},\ \Eprint
  {https://arxiv.org/abs/2308.06230} {arXiv:2308.06230 [hep-ex]} \BibitemShut
  {NoStop}%
\bibitem [{\citenamefont {Czarnecki}\ \emph {et~al.}(2003)\citenamefont
  {Czarnecki}, \citenamefont {Marciano},\ and\ \citenamefont
  {Vainshtein}}]{Czarnecki:2002nt}%
  \BibitemOpen
  \bibfield  {author} {\bibinfo {author} {\bibfnamefont {A.}~\bibnamefont
  {Czarnecki}}, \bibinfo {author} {\bibfnamefont {W.~J.}\ \bibnamefont
  {Marciano}},\ and\ \bibinfo {author} {\bibfnamefont {A.}~\bibnamefont
  {Vainshtein}},\ }\bibfield  {title} {\bibinfo {title} {{Refinements in
  electroweak contributions to the muon anomalous magnetic moment}},\ }\href
  {https://doi.org/10.1103/PhysRevD.67.073006} {\bibfield  {journal} {\bibinfo
  {journal} {Phys. Rev.}\ }\textbf {\bibinfo {volume} {D67}},\ \bibinfo {pages}
  {073006} (\bibinfo {year} {2003})},\ \bibinfo {note} {[Erratum: Phys. Rev.
  {\bf D73}, 119901 (2006)]},\ \Eprint {https://arxiv.org/abs/hep-ph/0212229}
  {arXiv:hep-ph/0212229 [hep-ph]} \BibitemShut {NoStop}%
\bibitem [{\citenamefont {Melnikov}\ and\ \citenamefont
  {Vainshtein}(2004)}]{Melnikov:2003xd}%
  \BibitemOpen
  \bibfield  {author} {\bibinfo {author} {\bibfnamefont {K.}~\bibnamefont
  {Melnikov}}\ and\ \bibinfo {author} {\bibfnamefont {A.}~\bibnamefont
  {Vainshtein}},\ }\bibfield  {title} {\bibinfo {title} {{Hadronic
  light-by-light scattering contribution to the muon anomalous magnetic moment
  revisited}},\ }\href {https://doi.org/10.1103/PhysRevD.70.113006} {\bibfield
  {journal} {\bibinfo  {journal} {Phys. Rev.}\ }\textbf {\bibinfo {volume}
  {D70}},\ \bibinfo {pages} {113006} (\bibinfo {year} {2004})},\ \Eprint
  {https://arxiv.org/abs/hep-ph/0312226} {arXiv:hep-ph/0312226 [hep-ph]}
  \BibitemShut {NoStop}%
\bibitem [{\citenamefont {Aoyama}\ \emph {et~al.}(2012)\citenamefont {Aoyama},
  \citenamefont {Hayakawa}, \citenamefont {Kinoshita},\ and\ \citenamefont
  {Nio}}]{Aoyama:2012wk}%
  \BibitemOpen
  \bibfield  {author} {\bibinfo {author} {\bibfnamefont {T.}~\bibnamefont
  {Aoyama}}, \bibinfo {author} {\bibfnamefont {M.}~\bibnamefont {Hayakawa}},
  \bibinfo {author} {\bibfnamefont {T.}~\bibnamefont {Kinoshita}},\ and\
  \bibinfo {author} {\bibfnamefont {M.}~\bibnamefont {Nio}},\ }\bibfield
  {title} {\bibinfo {title} {{Complete Tenth-Order QED Contribution to the Muon
  $g-2$}},\ }\href {https://doi.org/10.1103/PhysRevLett.109.111808} {\bibfield
  {journal} {\bibinfo  {journal} {Phys. Rev. Lett.}\ }\textbf {\bibinfo
  {volume} {109}},\ \bibinfo {pages} {111808} (\bibinfo {year} {2012})},\
  \Eprint {https://arxiv.org/abs/1205.5370} {arXiv:1205.5370 [hep-ph]}
  \BibitemShut {NoStop}%
\bibitem [{\citenamefont {Gnendiger}\ \emph {et~al.}(2013)\citenamefont
  {Gnendiger}, \citenamefont {St{\"o}ckinger},\ and\ \citenamefont
  {St{\"o}ckinger-Kim}}]{Gnendiger:2013pva}%
  \BibitemOpen
  \bibfield  {author} {\bibinfo {author} {\bibfnamefont {C.}~\bibnamefont
  {Gnendiger}}, \bibinfo {author} {\bibfnamefont {D.}~\bibnamefont
  {St{\"o}ckinger}},\ and\ \bibinfo {author} {\bibfnamefont {H.}~\bibnamefont
  {St{\"o}ckinger-Kim}},\ }\bibfield  {title} {\bibinfo {title} {{The
  electroweak contributions to $(g-2)_\mu$ after the Higgs boson mass
  measurement}},\ }\href {https://doi.org/10.1103/PhysRevD.88.053005}
  {\bibfield  {journal} {\bibinfo  {journal} {Phys. Rev.}\ }\textbf {\bibinfo
  {volume} {D88}},\ \bibinfo {pages} {053005} (\bibinfo {year} {2013})},\
  \Eprint {https://arxiv.org/abs/1306.5546} {arXiv:1306.5546 [hep-ph]}
  \BibitemShut {NoStop}%
\bibitem [{\citenamefont {Kurz}\ \emph {et~al.}(2014)\citenamefont {Kurz},
  \citenamefont {Liu}, \citenamefont {Marquard},\ and\ \citenamefont
  {Steinhauser}}]{Kurz:2014wya}%
  \BibitemOpen
  \bibfield  {author} {\bibinfo {author} {\bibfnamefont {A.}~\bibnamefont
  {Kurz}}, \bibinfo {author} {\bibfnamefont {T.}~\bibnamefont {Liu}}, \bibinfo
  {author} {\bibfnamefont {P.}~\bibnamefont {Marquard}},\ and\ \bibinfo
  {author} {\bibfnamefont {M.}~\bibnamefont {Steinhauser}},\ }\bibfield
  {title} {\bibinfo {title} {{Hadronic contribution to the muon anomalous
  magnetic moment to next-to-next-to-leading order}},\ }\href
  {https://doi.org/10.1016/j.physletb.2014.05.043} {\bibfield  {journal}
  {\bibinfo  {journal} {Phys. Lett.}\ }\textbf {\bibinfo {volume} {B734}},\
  \bibinfo {pages} {144} (\bibinfo {year} {2014})},\ \Eprint
  {https://arxiv.org/abs/1403.6400} {arXiv:1403.6400 [hep-ph]} \BibitemShut
  {NoStop}%
\bibitem [{\citenamefont {Colangelo}\ \emph {et~al.}(2014)\citenamefont
  {Colangelo}, \citenamefont {Hoferichter}, \citenamefont {Nyffeler},
  \citenamefont {Passera},\ and\ \citenamefont {Stoffer}}]{Colangelo:2014qya}%
  \BibitemOpen
  \bibfield  {author} {\bibinfo {author} {\bibfnamefont {G.}~\bibnamefont
  {Colangelo}}, \bibinfo {author} {\bibfnamefont {M.}~\bibnamefont
  {Hoferichter}}, \bibinfo {author} {\bibfnamefont {A.}~\bibnamefont
  {Nyffeler}}, \bibinfo {author} {\bibfnamefont {M.}~\bibnamefont {Passera}},\
  and\ \bibinfo {author} {\bibfnamefont {P.}~\bibnamefont {Stoffer}},\
  }\bibfield  {title} {\bibinfo {title} {{Remarks on higher-order hadronic
  corrections to the muon $g-2$}},\ }\href
  {https://doi.org/10.1016/j.physletb.2014.06.012} {\bibfield  {journal}
  {\bibinfo  {journal} {Phys. Lett.}\ }\textbf {\bibinfo {volume} {B735}},\
  \bibinfo {pages} {90} (\bibinfo {year} {2014})},\ \Eprint
  {https://arxiv.org/abs/1403.7512} {arXiv:1403.7512 [hep-ph]} \BibitemShut
  {NoStop}%
\bibitem [{\citenamefont {Davier}\ \emph {et~al.}(2017)\citenamefont {Davier},
  \citenamefont {Hoecker}, \citenamefont {Malaescu},\ and\ \citenamefont
  {Zhang}}]{Davier:2017zfy}%
  \BibitemOpen
  \bibfield  {author} {\bibinfo {author} {\bibfnamefont {M.}~\bibnamefont
  {Davier}}, \bibinfo {author} {\bibfnamefont {A.}~\bibnamefont {Hoecker}},
  \bibinfo {author} {\bibfnamefont {B.}~\bibnamefont {Malaescu}},\ and\
  \bibinfo {author} {\bibfnamefont {Z.}~\bibnamefont {Zhang}},\ }\bibfield
  {title} {\bibinfo {title} {{Reevaluation of the hadronic vacuum polarisation
  contributions to the Standard Model predictions of the muon $g-2$ and
  ${\alpha (m_Z^2)}$ using newest hadronic cross-section data}},\ }\href
  {https://doi.org/10.1140/epjc/s10052-017-5161-6} {\bibfield  {journal}
  {\bibinfo  {journal} {Eur. Phys. J.}\ }\textbf {\bibinfo {volume} {C77}},\
  \bibinfo {pages} {827} (\bibinfo {year} {2017})},\ \Eprint
  {https://arxiv.org/abs/1706.09436} {arXiv:1706.09436 [hep-ph]} \BibitemShut
  {NoStop}%
\bibitem [{\citenamefont {Masjuan}\ and\ \citenamefont
  {S{\'a}nchez-Puertas}(2017)}]{Masjuan:2017tvw}%
  \BibitemOpen
  \bibfield  {author} {\bibinfo {author} {\bibfnamefont {P.}~\bibnamefont
  {Masjuan}}\ and\ \bibinfo {author} {\bibfnamefont {P.}~\bibnamefont
  {S{\'a}nchez-Puertas}},\ }\bibfield  {title} {\bibinfo {title}
  {{Pseudoscalar-pole contribution to the $(g_{\mu}-2)$: a rational
  approach}},\ }\href {https://doi.org/10.1103/PhysRevD.95.054026} {\bibfield
  {journal} {\bibinfo  {journal} {Phys. Rev.}\ }\textbf {\bibinfo {volume}
  {D95}},\ \bibinfo {pages} {054026} (\bibinfo {year} {2017})},\ \Eprint
  {https://arxiv.org/abs/1701.05829} {arXiv:1701.05829 [hep-ph]} \BibitemShut
  {NoStop}%
\bibitem [{\citenamefont {Colangelo}\ \emph {et~al.}(2017)\citenamefont
  {Colangelo}, \citenamefont {Hoferichter}, \citenamefont {Procura},\ and\
  \citenamefont {Stoffer}}]{Colangelo:2017fiz}%
  \BibitemOpen
  \bibfield  {author} {\bibinfo {author} {\bibfnamefont {G.}~\bibnamefont
  {Colangelo}}, \bibinfo {author} {\bibfnamefont {M.}~\bibnamefont
  {Hoferichter}}, \bibinfo {author} {\bibfnamefont {M.}~\bibnamefont
  {Procura}},\ and\ \bibinfo {author} {\bibfnamefont {P.}~\bibnamefont
  {Stoffer}},\ }\bibfield  {title} {\bibinfo {title} {{Dispersion relation for
  hadronic light-by-light scattering: two-pion contributions}},\ }\href
  {https://doi.org/10.1007/JHEP04(2017)161} {\bibfield  {journal} {\bibinfo
  {journal} {JHEP}\ }\textbf {\bibinfo {volume} {04}},\ \bibinfo {pages}
  {161}},\ \Eprint {https://arxiv.org/abs/1702.07347} {arXiv:1702.07347
  [hep-ph]} \BibitemShut {NoStop}%
\bibitem [{\citenamefont {Keshavarzi}\ \emph {et~al.}(2018)\citenamefont
  {Keshavarzi}, \citenamefont {Nomura},\ and\ \citenamefont
  {Teubner}}]{Keshavarzi:2018mgv}%
  \BibitemOpen
  \bibfield  {author} {\bibinfo {author} {\bibfnamefont {A.}~\bibnamefont
  {Keshavarzi}}, \bibinfo {author} {\bibfnamefont {D.}~\bibnamefont {Nomura}},\
  and\ \bibinfo {author} {\bibfnamefont {T.}~\bibnamefont {Teubner}},\
  }\bibfield  {title} {\bibinfo {title} {{Muon $g-2$ and $\alpha(M_Z^2)$: a new
  data-based analysis}},\ }\href {https://doi.org/10.1103/PhysRevD.97.114025}
  {\bibfield  {journal} {\bibinfo  {journal} {Phys. Rev.}\ }\textbf {\bibinfo
  {volume} {D97}},\ \bibinfo {pages} {114025} (\bibinfo {year} {2018})},\
  \Eprint {https://arxiv.org/abs/1802.02995} {arXiv:1802.02995 [hep-ph]}
  \BibitemShut {NoStop}%
\bibitem [{\citenamefont {Colangelo}\ \emph {et~al.}(2019)\citenamefont
  {Colangelo}, \citenamefont {Hoferichter},\ and\ \citenamefont
  {Stoffer}}]{Colangelo:2018mtw}%
  \BibitemOpen
  \bibfield  {author} {\bibinfo {author} {\bibfnamefont {G.}~\bibnamefont
  {Colangelo}}, \bibinfo {author} {\bibfnamefont {M.}~\bibnamefont
  {Hoferichter}},\ and\ \bibinfo {author} {\bibfnamefont {P.}~\bibnamefont
  {Stoffer}},\ }\bibfield  {title} {\bibinfo {title} {{Two-pion contribution to
  hadronic vacuum polarization}},\ }\href
  {https://doi.org/10.1007/JHEP02(2019)006} {\bibfield  {journal} {\bibinfo
  {journal} {JHEP}\ }\textbf {\bibinfo {volume} {02}},\ \bibinfo {pages}
  {006}},\ \Eprint {https://arxiv.org/abs/1810.00007} {arXiv:1810.00007
  [hep-ph]} \BibitemShut {NoStop}%
\bibitem [{\citenamefont {Hoferichter}\ \emph {et~al.}(2018)\citenamefont
  {Hoferichter}, \citenamefont {Hoid}, \citenamefont {Kubis}, \citenamefont
  {Leupold},\ and\ \citenamefont {Schneider}}]{Hoferichter:2018kwz}%
  \BibitemOpen
  \bibfield  {author} {\bibinfo {author} {\bibfnamefont {M.}~\bibnamefont
  {Hoferichter}}, \bibinfo {author} {\bibfnamefont {B.-L.}\ \bibnamefont
  {Hoid}}, \bibinfo {author} {\bibfnamefont {B.}~\bibnamefont {Kubis}},
  \bibinfo {author} {\bibfnamefont {S.}~\bibnamefont {Leupold}},\ and\ \bibinfo
  {author} {\bibfnamefont {S.~P.}\ \bibnamefont {Schneider}},\ }\bibfield
  {title} {\bibinfo {title} {{Dispersion relation for hadronic light-by-light
  scattering: pion pole}},\ }\href {https://doi.org/10.1007/JHEP10(2018)141}
  {\bibfield  {journal} {\bibinfo  {journal} {JHEP}\ }\textbf {\bibinfo
  {volume} {10}},\ \bibinfo {pages} {141}},\ \Eprint
  {https://arxiv.org/abs/1808.04823} {arXiv:1808.04823 [hep-ph]} \BibitemShut
  {NoStop}%
\bibitem [{\citenamefont {G{\'e}rardin}\ \emph {et~al.}(2019)\citenamefont
  {G{\'e}rardin}, \citenamefont {Meyer},\ and\ \citenamefont
  {Nyffeler}}]{Gerardin:2019vio}%
  \BibitemOpen
  \bibfield  {author} {\bibinfo {author} {\bibfnamefont {A.}~\bibnamefont
  {G{\'e}rardin}}, \bibinfo {author} {\bibfnamefont {H.~B.}\ \bibnamefont
  {Meyer}},\ and\ \bibinfo {author} {\bibfnamefont {A.}~\bibnamefont
  {Nyffeler}},\ }\bibfield  {title} {\bibinfo {title} {{Lattice calculation of
  the pion transition form factor with $N_f=2+1$ Wilson quarks}},\ }\href
  {https://doi.org/10.1103/PhysRevD.100.034520} {\bibfield  {journal} {\bibinfo
   {journal} {Phys. Rev.}\ }\textbf {\bibinfo {volume} {D100}},\ \bibinfo
  {pages} {034520} (\bibinfo {year} {2019})},\ \Eprint
  {https://arxiv.org/abs/1903.09471} {arXiv:1903.09471 [hep-lat]} \BibitemShut
  {NoStop}%
\bibitem [{\citenamefont {Bijnens}\ \emph {et~al.}(2019)\citenamefont
  {Bijnens}, \citenamefont {Hermansson-Truedsson},\ and\ \citenamefont
  {Rodr{\'i}guez-S{\'a}nchez}}]{Bijnens:2019ghy}%
  \BibitemOpen
  \bibfield  {author} {\bibinfo {author} {\bibfnamefont {J.}~\bibnamefont
  {Bijnens}}, \bibinfo {author} {\bibfnamefont {N.}~\bibnamefont
  {Hermansson-Truedsson}},\ and\ \bibinfo {author} {\bibfnamefont
  {A.}~\bibnamefont {Rodr{\'i}guez-S{\'a}nchez}},\ }\bibfield  {title}
  {\bibinfo {title} {{Short-distance constraints for the HLbL contribution to
  the muon anomalous magnetic moment}},\ }\href
  {https://doi.org/10.1016/j.physletb.2019.134994} {\bibfield  {journal}
  {\bibinfo  {journal} {Phys. Lett.}\ }\textbf {\bibinfo {volume} {B798}},\
  \bibinfo {pages} {134994} (\bibinfo {year} {2019})},\ \Eprint
  {https://arxiv.org/abs/1908.03331} {arXiv:1908.03331 [hep-ph]} \BibitemShut
  {NoStop}%
\bibitem [{\citenamefont {Colangelo}\ \emph {et~al.}(2020)\citenamefont
  {Colangelo}, \citenamefont {Hagelstein}, \citenamefont {Hoferichter},
  \citenamefont {Laub},\ and\ \citenamefont {Stoffer}}]{Colangelo:2019uex}%
  \BibitemOpen
  \bibfield  {author} {\bibinfo {author} {\bibfnamefont {G.}~\bibnamefont
  {Colangelo}}, \bibinfo {author} {\bibfnamefont {F.}~\bibnamefont
  {Hagelstein}}, \bibinfo {author} {\bibfnamefont {M.}~\bibnamefont
  {Hoferichter}}, \bibinfo {author} {\bibfnamefont {L.}~\bibnamefont {Laub}},\
  and\ \bibinfo {author} {\bibfnamefont {P.}~\bibnamefont {Stoffer}},\
  }\bibfield  {title} {\bibinfo {title} {{Longitudinal short-distance
  constraints for the hadronic light-by-light contribution to $(g-2)_\mu$ with
  large-$N_c$ Regge models}},\ }\href {https://doi.org/10.1007/JHEP03(2020)101}
  {\bibfield  {journal} {\bibinfo  {journal} {JHEP}\ }\textbf {\bibinfo
  {volume} {03}},\ \bibinfo {pages} {101}},\ \Eprint
  {https://arxiv.org/abs/1910.13432} {arXiv:1910.13432 [hep-ph]} \BibitemShut
  {NoStop}%
\bibitem [{\citenamefont {Blum}\ \emph {et~al.}(2020)\citenamefont {Blum},
  \citenamefont {Christ}, \citenamefont {Hayakawa}, \citenamefont {Izubuchi},
  \citenamefont {Jin}, \citenamefont {Jung},\ and\ \citenamefont
  {Lehner}}]{Blum:2019ugy}%
  \BibitemOpen
  \bibfield  {author} {\bibinfo {author} {\bibfnamefont {T.}~\bibnamefont
  {Blum}}, \bibinfo {author} {\bibfnamefont {N.}~\bibnamefont {Christ}},
  \bibinfo {author} {\bibfnamefont {M.}~\bibnamefont {Hayakawa}}, \bibinfo
  {author} {\bibfnamefont {T.}~\bibnamefont {Izubuchi}}, \bibinfo {author}
  {\bibfnamefont {L.}~\bibnamefont {Jin}}, \bibinfo {author} {\bibfnamefont
  {C.}~\bibnamefont {Jung}},\ and\ \bibinfo {author} {\bibfnamefont
  {C.}~\bibnamefont {Lehner}},\ }\bibfield  {title} {\bibinfo {title} {{The
  hadronic light-by-light scattering contribution to the muon anomalous
  magnetic moment from lattice QCD}},\ }\href
  {https://doi.org/10.1103/PhysRevLett.124.132002} {\bibfield  {journal}
  {\bibinfo  {journal} {Phys. Rev. Lett.}\ }\textbf {\bibinfo {volume} {124}},\
  \bibinfo {pages} {132002} (\bibinfo {year} {2020})},\ \Eprint
  {https://arxiv.org/abs/1911.08123} {arXiv:1911.08123 [hep-lat]} \BibitemShut
  {NoStop}%
\bibitem [{\citenamefont {Hoferichter}\ \emph {et~al.}(2019)\citenamefont
  {Hoferichter}, \citenamefont {Hoid},\ and\ \citenamefont
  {Kubis}}]{Hoferichter:2019mqg}%
  \BibitemOpen
  \bibfield  {author} {\bibinfo {author} {\bibfnamefont {M.}~\bibnamefont
  {Hoferichter}}, \bibinfo {author} {\bibfnamefont {B.-L.}\ \bibnamefont
  {Hoid}},\ and\ \bibinfo {author} {\bibfnamefont {B.}~\bibnamefont {Kubis}},\
  }\bibfield  {title} {\bibinfo {title} {{Three-pion contribution to hadronic
  vacuum polarization}},\ }\href {https://doi.org/10.1007/JHEP08(2019)137}
  {\bibfield  {journal} {\bibinfo  {journal} {JHEP}\ }\textbf {\bibinfo
  {volume} {08}},\ \bibinfo {pages} {137}},\ \Eprint
  {https://arxiv.org/abs/1907.01556} {arXiv:1907.01556 [hep-ph]} \BibitemShut
  {NoStop}%
\bibitem [{\citenamefont {Aoyama}\ \emph {et~al.}(2019)\citenamefont {Aoyama},
  \citenamefont {Kinoshita},\ and\ \citenamefont {Nio}}]{Aoyama:2019ryr}%
  \BibitemOpen
  \bibfield  {author} {\bibinfo {author} {\bibfnamefont {T.}~\bibnamefont
  {Aoyama}}, \bibinfo {author} {\bibfnamefont {T.}~\bibnamefont {Kinoshita}},\
  and\ \bibinfo {author} {\bibfnamefont {M.}~\bibnamefont {Nio}},\ }\bibfield
  {title} {\bibinfo {title} {{Theory of the Anomalous Magnetic Moment of the
  Electron}},\ }\href {https://doi.org/10.3390/atoms7010028} {\bibfield
  {journal} {\bibinfo  {journal} {Atoms}\ }\textbf {\bibinfo {volume} {7}},\
  \bibinfo {pages} {28} (\bibinfo {year} {2019})}\BibitemShut {NoStop}%
\bibitem [{\citenamefont {Davier}\ \emph {et~al.}(2020)\citenamefont {Davier},
  \citenamefont {Hoecker}, \citenamefont {Malaescu},\ and\ \citenamefont
  {Zhang}}]{Davier:2019can}%
  \BibitemOpen
  \bibfield  {author} {\bibinfo {author} {\bibfnamefont {M.}~\bibnamefont
  {Davier}}, \bibinfo {author} {\bibfnamefont {A.}~\bibnamefont {Hoecker}},
  \bibinfo {author} {\bibfnamefont {B.}~\bibnamefont {Malaescu}},\ and\
  \bibinfo {author} {\bibfnamefont {Z.}~\bibnamefont {Zhang}},\ }\bibfield
  {title} {\bibinfo {title} {{A new evaluation of the hadronic vacuum
  polarisation contributions to the muon anomalous magnetic moment and to
  $\mathbf{\boldsymbol\alpha(m_Z^2)}$}},\ }\href
  {https://doi.org/10.1140/epjc/s10052-020-7792-2} {\bibfield  {journal}
  {\bibinfo  {journal} {Eur. Phys. J.}\ }\textbf {\bibinfo {volume} {C80}},\
  \bibinfo {pages} {241} (\bibinfo {year} {2020})},\ \bibinfo {note} {[Erratum:
  Eur. Phys. J. {\bf C80}, 410 (2020)]},\ \Eprint
  {https://arxiv.org/abs/1908.00921} {arXiv:1908.00921 [hep-ph]} \BibitemShut
  {NoStop}%
\bibitem [{\citenamefont {Keshavarzi}\ \emph {et~al.}(2020)\citenamefont
  {Keshavarzi}, \citenamefont {Nomura},\ and\ \citenamefont
  {Teubner}}]{Keshavarzi:2019abf}%
  \BibitemOpen
  \bibfield  {author} {\bibinfo {author} {\bibfnamefont {A.}~\bibnamefont
  {Keshavarzi}}, \bibinfo {author} {\bibfnamefont {D.}~\bibnamefont {Nomura}},\
  and\ \bibinfo {author} {\bibfnamefont {T.}~\bibnamefont {Teubner}},\
  }\bibfield  {title} {\bibinfo {title} {{The $g-2$ of charged leptons,
  $\alpha(M_Z^2)$ and the hyperfine splitting of muonium}},\ }\href
  {https://doi.org/10.1103/PhysRevD.101.014029} {\bibfield  {journal} {\bibinfo
   {journal} {Phys. Rev.}\ }\textbf {\bibinfo {volume} {D101}},\ \bibinfo
  {pages} {014029} (\bibinfo {year} {2020})},\ \Eprint
  {https://arxiv.org/abs/1911.00367} {arXiv:1911.00367 [hep-ph]} \BibitemShut
  {NoStop}%
\bibitem [{\citenamefont {Aoyama}\ \emph {et~al.}(2020)\citenamefont {Aoyama}
  \emph {et~al.}}]{Aoyama:2020ynm}%
  \BibitemOpen
  \bibfield  {author} {\bibinfo {author} {\bibfnamefont {T.}~\bibnamefont
  {Aoyama}} \emph {et~al.},\ }\bibfield  {title} {\bibinfo {title} {{The
  anomalous magnetic moment of the muon in the Standard Model}},\ }\href
  {https://doi.org/10.1016/j.physrep.2020.07.006} {\bibfield  {journal}
  {\bibinfo  {journal} {Phys. Rept.}\ }\textbf {\bibinfo {volume} {887}},\
  \bibinfo {pages} {1} (\bibinfo {year} {2020})},\ \Eprint
  {https://arxiv.org/abs/2006.04822} {arXiv:2006.04822 [hep-ph]} \BibitemShut
  {NoStop}%
\bibitem [{\citenamefont {Baek}\ \emph {et~al.}(2001)\citenamefont {Baek},
  \citenamefont {Deshpande}, \citenamefont {He},\ and\ \citenamefont
  {Ko}}]{Baek:2001kca}%
  \BibitemOpen
  \bibfield  {author} {\bibinfo {author} {\bibfnamefont {S.}~\bibnamefont
  {Baek}}, \bibinfo {author} {\bibfnamefont {N.~G.}\ \bibnamefont {Deshpande}},
  \bibinfo {author} {\bibfnamefont {X.~G.}\ \bibnamefont {He}},\ and\ \bibinfo
  {author} {\bibfnamefont {P.}~\bibnamefont {Ko}},\ }\bibfield  {title}
  {\bibinfo {title} {{Muon anomalous g-2 and gauged L(muon) - L(tau) models}},\
  }\href {https://doi.org/10.1103/PhysRevD.64.055006} {\bibfield  {journal}
  {\bibinfo  {journal} {Phys. Rev. D}\ }\textbf {\bibinfo {volume} {64}},\
  \bibinfo {pages} {055006} (\bibinfo {year} {2001})},\ \Eprint
  {https://arxiv.org/abs/hep-ph/0104141} {arXiv:hep-ph/0104141} \BibitemShut
  {NoStop}%
\bibitem [{\citenamefont {Ma}\ \emph {et~al.}(2002)\citenamefont {Ma},
  \citenamefont {Roy},\ and\ \citenamefont {Roy}}]{Ma:2001md}%
  \BibitemOpen
  \bibfield  {author} {\bibinfo {author} {\bibfnamefont {E.}~\bibnamefont
  {Ma}}, \bibinfo {author} {\bibfnamefont {D.~P.}\ \bibnamefont {Roy}},\ and\
  \bibinfo {author} {\bibfnamefont {S.}~\bibnamefont {Roy}},\ }\bibfield
  {title} {\bibinfo {title} {{Gauged L(mu) - L(tau) with large muon anomalous
  magnetic moment and the bimaximal mixing of neutrinos}},\ }\href
  {https://doi.org/10.1016/S0370-2693(01)01428-9} {\bibfield  {journal}
  {\bibinfo  {journal} {Phys. Lett. B}\ }\textbf {\bibinfo {volume} {525}},\
  \bibinfo {pages} {101} (\bibinfo {year} {2002})},\ \Eprint
  {https://arxiv.org/abs/hep-ph/0110146} {arXiv:hep-ph/0110146} \BibitemShut
  {NoStop}%
\bibitem [{\citenamefont {Banerjee}\ \emph {et~al.}(2021)\citenamefont
  {Banerjee}, \citenamefont {Dutta},\ and\ \citenamefont
  {Roy}}]{Banerjee:2020zvi}%
  \BibitemOpen
  \bibfield  {author} {\bibinfo {author} {\bibfnamefont {H.}~\bibnamefont
  {Banerjee}}, \bibinfo {author} {\bibfnamefont {B.}~\bibnamefont {Dutta}},\
  and\ \bibinfo {author} {\bibfnamefont {S.}~\bibnamefont {Roy}},\ }\bibfield
  {title} {\bibinfo {title} {{Supersymmetric gauged $U(1)_{L_\mu -L_\tau}$
  model for electron and muon $(g-2)$ anomaly}},\ }\href
  {https://doi.org/10.1007/JHEP03(2021)211} {\bibfield  {journal} {\bibinfo
  {journal} {JHEP}\ }\textbf {\bibinfo {volume} {03}},\ \bibinfo {pages}
  {211}},\ \Eprint {https://arxiv.org/abs/2011.05083} {arXiv:2011.05083
  [hep-ph]} \BibitemShut {NoStop}%
\bibitem [{\citenamefont {Wittig}(2023)}]{Wittig:2023pcl}%
  \BibitemOpen
  \bibfield  {author} {\bibinfo {author} {\bibfnamefont {H.}~\bibnamefont
  {Wittig}},\ }\bibfield  {title} {\bibinfo {title} {{Progress on $(g-2)_\mu$
  from Lattice QCD}},\ }\bibfield  {booktitle} {\emph {\bibinfo {booktitle}
  {{57th Rencontres de Moriond on Electroweak Interactions and Unified
  Theories}}},\ }\href@noop {} {\  (\bibinfo {year} {2023})},\ \Eprint
  {https://arxiv.org/abs/2306.04165} {arXiv:2306.04165 [hep-ph]} \BibitemShut
  {NoStop}%
\bibitem [{\citenamefont {Ignatov}\ \emph {et~al.}(2024)\citenamefont {Ignatov}
  \emph {et~al.}}]{CMD-3:2023alj}%
  \BibitemOpen
  \bibfield  {author} {\bibinfo {author} {\bibfnamefont {F.~V.}\ \bibnamefont
  {Ignatov}} \emph {et~al.} (\bibinfo {collaboration} {CMD-3}),\ }\bibfield
  {title} {\bibinfo {title} {{Measurement of the
  e+e-\textrightarrow{}\ensuremath{\pi}+\ensuremath{\pi}- cross section from
  threshold to 1.2~GeV with the CMD-3 detector}},\ }\href
  {https://doi.org/10.1103/PhysRevD.109.112002} {\bibfield  {journal} {\bibinfo
   {journal} {Phys. Rev. D}\ }\textbf {\bibinfo {volume} {109}},\ \bibinfo
  {pages} {112002} (\bibinfo {year} {2024})},\ \Eprint
  {https://arxiv.org/abs/2302.08834} {arXiv:2302.08834 [hep-ex]} \BibitemShut
  {NoStop}%
\bibitem [{\citenamefont {'t~Hooft}(1976{\natexlab{a}})}]{PhysRevLett.37.8}%
  \BibitemOpen
  \bibfield  {author} {\bibinfo {author} {\bibfnamefont {G.}~\bibnamefont
  {'t~Hooft}},\ }\bibfield  {title} {\bibinfo {title} {Symmetry breaking
  through bell-jackiw anomalies},\ }\href
  {https://doi.org/10.1103/PhysRevLett.37.8} {\bibfield  {journal} {\bibinfo
  {journal} {Phys. Rev. Lett.}\ }\textbf {\bibinfo {volume} {37}},\ \bibinfo
  {pages} {8} (\bibinfo {year} {1976}{\natexlab{a}})}\BibitemShut {NoStop}%
\bibitem [{\citenamefont {'t~Hooft}(1976{\natexlab{b}})}]{PhysRevD.14.3432}%
  \BibitemOpen
  \bibfield  {author} {\bibinfo {author} {\bibfnamefont {G.}~\bibnamefont
  {'t~Hooft}},\ }\bibfield  {title} {\bibinfo {title} {Computation of the
  quantum effects due to a four-dimensional pseudoparticle},\ }\href
  {https://doi.org/10.1103/PhysRevD.14.3432} {\bibfield  {journal} {\bibinfo
  {journal} {Phys. Rev. D}\ }\textbf {\bibinfo {volume} {14}},\ \bibinfo
  {pages} {3432} (\bibinfo {year} {1976}{\natexlab{b}})}\BibitemShut {NoStop}%
\bibitem [{\citenamefont {Baker}\ \emph {et~al.}(2006)\citenamefont {Baker},
  \citenamefont {Doyle}, \citenamefont {Geltenbort}, \citenamefont {Green},
  \citenamefont {van~der Grinten}, \citenamefont {Harris}, \citenamefont
  {Iaydjiev}, \citenamefont {Ivanov}, \citenamefont {May}, \citenamefont
  {Pendlebury}, \citenamefont {Richardson}, \citenamefont {Shiers},\ and\
  \citenamefont {Smith}}]{PhysRevLett.97.131801}%
  \BibitemOpen
  \bibfield  {author} {\bibinfo {author} {\bibfnamefont {C.~A.}\ \bibnamefont
  {Baker}}, \bibinfo {author} {\bibfnamefont {D.~D.}\ \bibnamefont {Doyle}},
  \bibinfo {author} {\bibfnamefont {P.}~\bibnamefont {Geltenbort}}, \bibinfo
  {author} {\bibfnamefont {K.}~\bibnamefont {Green}}, \bibinfo {author}
  {\bibfnamefont {M.~G.~D.}\ \bibnamefont {van~der Grinten}}, \bibinfo {author}
  {\bibfnamefont {P.~G.}\ \bibnamefont {Harris}}, \bibinfo {author}
  {\bibfnamefont {P.}~\bibnamefont {Iaydjiev}}, \bibinfo {author}
  {\bibfnamefont {S.~N.}\ \bibnamefont {Ivanov}}, \bibinfo {author}
  {\bibfnamefont {D.~J.~R.}\ \bibnamefont {May}}, \bibinfo {author}
  {\bibfnamefont {J.~M.}\ \bibnamefont {Pendlebury}}, \bibinfo {author}
  {\bibfnamefont {J.~D.}\ \bibnamefont {Richardson}}, \bibinfo {author}
  {\bibfnamefont {D.}~\bibnamefont {Shiers}},\ and\ \bibinfo {author}
  {\bibfnamefont {K.~F.}\ \bibnamefont {Smith}},\ }\bibfield  {title} {\bibinfo
  {title} {Improved experimental limit on the electric dipole moment of the
  neutron},\ }\href {https://doi.org/10.1103/PhysRevLett.97.131801} {\bibfield
  {journal} {\bibinfo  {journal} {Phys. Rev. Lett.}\ }\textbf {\bibinfo
  {volume} {97}},\ \bibinfo {pages} {131801} (\bibinfo {year}
  {2006})}\BibitemShut {NoStop}%
\bibitem [{\citenamefont {Peccei}\ and\ \citenamefont
  {Quinn}(1977{\natexlab{a}})}]{Peccei:1977ur}%
  \BibitemOpen
  \bibfield  {author} {\bibinfo {author} {\bibfnamefont {R.~D.}\ \bibnamefont
  {Peccei}}\ and\ \bibinfo {author} {\bibfnamefont {H.~R.}\ \bibnamefont
  {Quinn}},\ }\bibfield  {title} {\bibinfo {title} {{Constraints Imposed by CP
  Conservation in the Presence of Instantons}},\ }\href
  {https://doi.org/10.1103/PhysRevD.16.1791} {\bibfield  {journal} {\bibinfo
  {journal} {Phys. Rev. D}\ }\textbf {\bibinfo {volume} {16}},\ \bibinfo
  {pages} {1791} (\bibinfo {year} {1977}{\natexlab{a}})}\BibitemShut {NoStop}%
\bibitem [{\citenamefont {Peccei}\ and\ \citenamefont
  {Quinn}(1977{\natexlab{b}})}]{Peccei:1977hh}%
  \BibitemOpen
  \bibfield  {author} {\bibinfo {author} {\bibfnamefont {R.~D.}\ \bibnamefont
  {Peccei}}\ and\ \bibinfo {author} {\bibfnamefont {H.~R.}\ \bibnamefont
  {Quinn}},\ }\bibfield  {title} {\bibinfo {title} {{CP Conservation in the
  Presence of Instantons}},\ }\href
  {https://doi.org/10.1103/PhysRevLett.38.1440} {\bibfield  {journal} {\bibinfo
   {journal} {Phys. Rev. Lett.}\ }\textbf {\bibinfo {volume} {38}},\ \bibinfo
  {pages} {1440} (\bibinfo {year} {1977}{\natexlab{b}})}\BibitemShut {NoStop}%
\bibitem [{\citenamefont {Weinberg}(1978)}]{PhysRevLett.40.223}%
  \BibitemOpen
  \bibfield  {author} {\bibinfo {author} {\bibfnamefont {S.}~\bibnamefont
  {Weinberg}},\ }\bibfield  {title} {\bibinfo {title} {A new light boson?},\
  }\href {https://doi.org/10.1103/PhysRevLett.40.223} {\bibfield  {journal}
  {\bibinfo  {journal} {Phys. Rev. Lett.}\ }\textbf {\bibinfo {volume} {40}},\
  \bibinfo {pages} {223} (\bibinfo {year} {1978})}\BibitemShut {NoStop}%
\bibitem [{\citenamefont {Wilczek}(1978)}]{PhysRevLett.40.279}%
  \BibitemOpen
  \bibfield  {author} {\bibinfo {author} {\bibfnamefont {F.}~\bibnamefont
  {Wilczek}},\ }\bibfield  {title} {\bibinfo {title} {Problem of strong $p$ and
  $t$ invariance in the presence of instantons},\ }\href
  {https://doi.org/10.1103/PhysRevLett.40.279} {\bibfield  {journal} {\bibinfo
  {journal} {Phys. Rev. Lett.}\ }\textbf {\bibinfo {volume} {40}},\ \bibinfo
  {pages} {279} (\bibinfo {year} {1978})}\BibitemShut {NoStop}%
\bibitem [{\citenamefont {Dine}\ and\ \citenamefont
  {Fischler}(1983)}]{Dine:1982ah}%
  \BibitemOpen
  \bibfield  {author} {\bibinfo {author} {\bibfnamefont {M.}~\bibnamefont
  {Dine}}\ and\ \bibinfo {author} {\bibfnamefont {W.}~\bibnamefont
  {Fischler}},\ }\bibfield  {title} {\bibinfo {title} {{The Not So Harmless
  Axion}},\ }\href {https://doi.org/10.1016/0370-2693(83)90639-1} {\bibfield
  {journal} {\bibinfo  {journal} {Phys. Lett. B}\ }\textbf {\bibinfo {volume}
  {120}},\ \bibinfo {pages} {137} (\bibinfo {year} {1983})}\BibitemShut
  {NoStop}%
\bibitem [{\citenamefont {Preskill}\ \emph {et~al.}(1983)\citenamefont
  {Preskill}, \citenamefont {Wise},\ and\ \citenamefont
  {Wilczek}}]{PRESKILL1983127}%
  \BibitemOpen
  \bibfield  {author} {\bibinfo {author} {\bibfnamefont {J.}~\bibnamefont
  {Preskill}}, \bibinfo {author} {\bibfnamefont {M.~B.}\ \bibnamefont {Wise}},\
  and\ \bibinfo {author} {\bibfnamefont {F.}~\bibnamefont {Wilczek}},\
  }\bibfield  {title} {\bibinfo {title} {Cosmology of the invisible axion},\
  }\href {https://doi.org/https://doi.org/10.1016/0370-2693(83)90637-8}
  {\bibfield  {journal} {\bibinfo  {journal} {Physics Letters B}\ }\textbf
  {\bibinfo {volume} {120}},\ \bibinfo {pages} {127} (\bibinfo {year}
  {1983})}\BibitemShut {NoStop}%
\bibitem [{\citenamefont {Abbott}\ and\ \citenamefont
  {Sikivie}(1983)}]{ABBOTT1983133}%
  \BibitemOpen
  \bibfield  {author} {\bibinfo {author} {\bibfnamefont {L.}~\bibnamefont
  {Abbott}}\ and\ \bibinfo {author} {\bibfnamefont {P.}~\bibnamefont
  {Sikivie}},\ }\bibfield  {title} {\bibinfo {title} {A cosmological bound on
  the invisible axion},\ }\href
  {https://doi.org/https://doi.org/10.1016/0370-2693(83)90638-X} {\bibfield
  {journal} {\bibinfo  {journal} {Physics Letters B}\ }\textbf {\bibinfo
  {volume} {120}},\ \bibinfo {pages} {133} (\bibinfo {year}
  {1983})}\BibitemShut {NoStop}%
\bibitem [{\citenamefont {Marsh}(2016)}]{MARSH20161}%
  \BibitemOpen
  \bibfield  {author} {\bibinfo {author} {\bibfnamefont {D.~J.}\ \bibnamefont
  {Marsh}},\ }\bibfield  {title} {\bibinfo {title} {Axion cosmology},\ }\href
  {https://doi.org/https://doi.org/10.1016/j.physrep.2016.06.005} {\bibfield
  {journal} {\bibinfo  {journal} {Physics Reports}\ }\textbf {\bibinfo {volume}
  {643}},\ \bibinfo {pages} {1} (\bibinfo {year} {2016})},\ \bibinfo {note}
  {axion cosmology}\BibitemShut {NoStop}%
\bibitem [{\citenamefont {{Di Luzio}}\ \emph {et~al.}(2020)\citenamefont {{Di
  Luzio}}, \citenamefont {Giannotti}, \citenamefont {Nardi},\ and\
  \citenamefont {Visinelli}}]{DILUZIO20201}%
  \BibitemOpen
  \bibfield  {author} {\bibinfo {author} {\bibfnamefont {L.}~\bibnamefont {{Di
  Luzio}}}, \bibinfo {author} {\bibfnamefont {M.}~\bibnamefont {Giannotti}},
  \bibinfo {author} {\bibfnamefont {E.}~\bibnamefont {Nardi}},\ and\ \bibinfo
  {author} {\bibfnamefont {L.}~\bibnamefont {Visinelli}},\ }\bibfield  {title}
  {\bibinfo {title} {The landscape of qcd axion models},\ }\href
  {https://doi.org/https://doi.org/10.1016/j.physrep.2020.06.002} {\bibfield
  {journal} {\bibinfo  {journal} {Physics Reports}\ }\textbf {\bibinfo {volume}
  {870}},\ \bibinfo {pages} {1} (\bibinfo {year} {2020})},\ \bibinfo {note}
  {the landscape of QCD axion models}\BibitemShut {NoStop}%
\bibitem [{\citenamefont {Arvanitaki}\ \emph {et~al.}(2010)\citenamefont
  {Arvanitaki}, \citenamefont {Dimopoulos}, \citenamefont {Dubovsky},
  \citenamefont {Kaloper},\ and\ \citenamefont
  {March-Russell}}]{Arvanitaki:2009fg}%
  \BibitemOpen
  \bibfield  {author} {\bibinfo {author} {\bibfnamefont {A.}~\bibnamefont
  {Arvanitaki}}, \bibinfo {author} {\bibfnamefont {S.}~\bibnamefont
  {Dimopoulos}}, \bibinfo {author} {\bibfnamefont {S.}~\bibnamefont
  {Dubovsky}}, \bibinfo {author} {\bibfnamefont {N.}~\bibnamefont {Kaloper}},\
  and\ \bibinfo {author} {\bibfnamefont {J.}~\bibnamefont {March-Russell}},\
  }\bibfield  {title} {\bibinfo {title} {{String Axiverse}},\ }\href
  {https://doi.org/10.1103/PhysRevD.81.123530} {\bibfield  {journal} {\bibinfo
  {journal} {Phys. Rev. D}\ }\textbf {\bibinfo {volume} {81}},\ \bibinfo
  {pages} {123530} (\bibinfo {year} {2010})},\ \Eprint
  {https://arxiv.org/abs/0905.4720} {arXiv:0905.4720 [hep-th]} \BibitemShut
  {NoStop}%
\bibitem [{\citenamefont {Cicoli}\ \emph {et~al.}(2012)\citenamefont {Cicoli},
  \citenamefont {Goodsell},\ and\ \citenamefont {Ringwald}}]{Cicoli:2012sz}%
  \BibitemOpen
  \bibfield  {author} {\bibinfo {author} {\bibfnamefont {M.}~\bibnamefont
  {Cicoli}}, \bibinfo {author} {\bibfnamefont {M.}~\bibnamefont {Goodsell}},\
  and\ \bibinfo {author} {\bibfnamefont {A.}~\bibnamefont {Ringwald}},\
  }\bibfield  {title} {\bibinfo {title} {{The type IIB string axiverse and its
  low-energy phenomenology}},\ }\href {https://doi.org/10.1007/JHEP10(2012)146}
  {\bibfield  {journal} {\bibinfo  {journal} {JHEP}\ }\textbf {\bibinfo
  {volume} {10}},\ \bibinfo {pages} {146}},\ \Eprint
  {https://arxiv.org/abs/1206.0819} {arXiv:1206.0819 [hep-th]} \BibitemShut
  {NoStop}%
\bibitem [{\citenamefont {Visinelli}\ and\ \citenamefont
  {Vagnozzi}(2019)}]{Visinelli:2018utg}%
  \BibitemOpen
  \bibfield  {author} {\bibinfo {author} {\bibfnamefont {L.}~\bibnamefont
  {Visinelli}}\ and\ \bibinfo {author} {\bibfnamefont {S.}~\bibnamefont
  {Vagnozzi}},\ }\bibfield  {title} {\bibinfo {title} {{Cosmological window
  onto the string axiverse and the supersymmetry breaking scale}},\ }\href
  {https://doi.org/10.1103/PhysRevD.99.063517} {\bibfield  {journal} {\bibinfo
  {journal} {Phys. Rev. D}\ }\textbf {\bibinfo {volume} {99}},\ \bibinfo
  {pages} {063517} (\bibinfo {year} {2019})},\ \Eprint
  {https://arxiv.org/abs/1809.06382} {arXiv:1809.06382 [hep-ph]} \BibitemShut
  {NoStop}%
\bibitem [{\citenamefont {Bauer}\ \emph {et~al.}(2017)\citenamefont {Bauer},
  \citenamefont {Neubert},\ and\ \citenamefont {Thamm}}]{Bauer:2017ris}%
  \BibitemOpen
  \bibfield  {author} {\bibinfo {author} {\bibfnamefont {M.}~\bibnamefont
  {Bauer}}, \bibinfo {author} {\bibfnamefont {M.}~\bibnamefont {Neubert}},\
  and\ \bibinfo {author} {\bibfnamefont {A.}~\bibnamefont {Thamm}},\ }\bibfield
   {title} {\bibinfo {title} {{Collider Probes of Axion-Like Particles}},\
  }\href {https://doi.org/10.1007/JHEP12(2017)044} {\bibfield  {journal}
  {\bibinfo  {journal} {JHEP}\ }\textbf {\bibinfo {volume} {12}},\ \bibinfo
  {pages} {044}},\ \Eprint {https://arxiv.org/abs/1708.00443} {arXiv:1708.00443
  [hep-ph]} \BibitemShut {NoStop}%
\bibitem [{\citenamefont {Bauer}\ \emph {et~al.}(2020)\citenamefont {Bauer},
  \citenamefont {Neubert}, \citenamefont {Renner}, \citenamefont {Schnubel},\
  and\ \citenamefont {Thamm}}]{Bauer:2019gfk}%
  \BibitemOpen
  \bibfield  {author} {\bibinfo {author} {\bibfnamefont {M.}~\bibnamefont
  {Bauer}}, \bibinfo {author} {\bibfnamefont {M.}~\bibnamefont {Neubert}},
  \bibinfo {author} {\bibfnamefont {S.}~\bibnamefont {Renner}}, \bibinfo
  {author} {\bibfnamefont {M.}~\bibnamefont {Schnubel}},\ and\ \bibinfo
  {author} {\bibfnamefont {A.}~\bibnamefont {Thamm}},\ }\bibfield  {title}
  {\bibinfo {title} {{Axionlike Particles, Lepton-Flavor Violation, and a New
  Explanation of $a_\mu$ and $a_e$}},\ }\href
  {https://doi.org/10.1103/PhysRevLett.124.211803} {\bibfield  {journal}
  {\bibinfo  {journal} {Phys. Rev. Lett.}\ }\textbf {\bibinfo {volume} {124}},\
  \bibinfo {pages} {211803} (\bibinfo {year} {2020})},\ \Eprint
  {https://arxiv.org/abs/1908.00008} {arXiv:1908.00008 [hep-ph]} \BibitemShut
  {NoStop}%
\bibitem [{\citenamefont {Calibbi}\ \emph {et~al.}(2021)\citenamefont
  {Calibbi}, \citenamefont {Redigolo}, \citenamefont {Ziegler},\ and\
  \citenamefont {Zupan}}]{Calibbi:2020jvd}%
  \BibitemOpen
  \bibfield  {author} {\bibinfo {author} {\bibfnamefont {L.}~\bibnamefont
  {Calibbi}}, \bibinfo {author} {\bibfnamefont {D.}~\bibnamefont {Redigolo}},
  \bibinfo {author} {\bibfnamefont {R.}~\bibnamefont {Ziegler}},\ and\ \bibinfo
  {author} {\bibfnamefont {J.}~\bibnamefont {Zupan}},\ }\bibfield  {title}
  {\bibinfo {title} {{Looking forward to Lepton-flavor-violating ALPs}},\
  }\href {https://doi.org/10.1007/JHEP09(2021)173} {\bibfield  {journal}
  {\bibinfo  {journal} {JHEP}\ }\textbf {\bibinfo {volume} {09}},\ \bibinfo
  {pages} {173}},\ \Eprint {https://arxiv.org/abs/2006.04795} {arXiv:2006.04795
  [hep-ph]} \BibitemShut {NoStop}%
\bibitem [{\citenamefont {Bauer}\ \emph {et~al.}(2021)\citenamefont {Bauer},
  \citenamefont {Neubert}, \citenamefont {Renner}, \citenamefont {Schnubel},\
  and\ \citenamefont {Thamm}}]{Bauer:2021mvw}%
  \BibitemOpen
  \bibfield  {author} {\bibinfo {author} {\bibfnamefont {M.}~\bibnamefont
  {Bauer}}, \bibinfo {author} {\bibfnamefont {M.}~\bibnamefont {Neubert}},
  \bibinfo {author} {\bibfnamefont {S.}~\bibnamefont {Renner}}, \bibinfo
  {author} {\bibfnamefont {M.}~\bibnamefont {Schnubel}},\ and\ \bibinfo
  {author} {\bibfnamefont {A.}~\bibnamefont {Thamm}},\ }\bibfield  {title}
  {\bibinfo {title} {{Flavor probes of axion-like particles}},\ }\href@noop {}
  {\  (\bibinfo {year} {2021})},\ \Eprint {https://arxiv.org/abs/2110.10698}
  {arXiv:2110.10698 [hep-ph]} \BibitemShut {NoStop}%
\bibitem [{\citenamefont {Bertholet}\ \emph {et~al.}(2022)\citenamefont
  {Bertholet}, \citenamefont {Chakraborty}, \citenamefont {Loladze},
  \citenamefont {Okui}, \citenamefont {Soffer},\ and\ \citenamefont
  {Tobioka}}]{Bertholet:2021hjl}%
  \BibitemOpen
  \bibfield  {author} {\bibinfo {author} {\bibfnamefont {E.}~\bibnamefont
  {Bertholet}}, \bibinfo {author} {\bibfnamefont {S.}~\bibnamefont
  {Chakraborty}}, \bibinfo {author} {\bibfnamefont {V.}~\bibnamefont
  {Loladze}}, \bibinfo {author} {\bibfnamefont {T.}~\bibnamefont {Okui}},
  \bibinfo {author} {\bibfnamefont {A.}~\bibnamefont {Soffer}},\ and\ \bibinfo
  {author} {\bibfnamefont {K.}~\bibnamefont {Tobioka}},\ }\bibfield  {title}
  {\bibinfo {title} {{Heavy QCD axion at Belle II: Displaced and prompt
  signals}},\ }\href {https://doi.org/10.1103/PhysRevD.105.L071701} {\bibfield
  {journal} {\bibinfo  {journal} {Phys. Rev. D}\ }\textbf {\bibinfo {volume}
  {105}},\ \bibinfo {pages} {L071701} (\bibinfo {year} {2022})},\ \Eprint
  {https://arxiv.org/abs/2108.10331} {arXiv:2108.10331 [hep-ph]} \BibitemShut
  {NoStop}%
\bibitem [{\citenamefont {Chakraborty}\ \emph {et~al.}(2021)\citenamefont
  {Chakraborty}, \citenamefont {Kraus}, \citenamefont {Loladze}, \citenamefont
  {Okui},\ and\ \citenamefont {Tobioka}}]{Chakraborty:2021wda}%
  \BibitemOpen
  \bibfield  {author} {\bibinfo {author} {\bibfnamefont {S.}~\bibnamefont
  {Chakraborty}}, \bibinfo {author} {\bibfnamefont {M.}~\bibnamefont {Kraus}},
  \bibinfo {author} {\bibfnamefont {V.}~\bibnamefont {Loladze}}, \bibinfo
  {author} {\bibfnamefont {T.}~\bibnamefont {Okui}},\ and\ \bibinfo {author}
  {\bibfnamefont {K.}~\bibnamefont {Tobioka}},\ }\bibfield  {title} {\bibinfo
  {title} {{Heavy QCD axion in b\textrightarrow{}s transition: Enhanced limits
  and projections}},\ }\href {https://doi.org/10.1103/PhysRevD.104.055036}
  {\bibfield  {journal} {\bibinfo  {journal} {Phys. Rev. D}\ }\textbf {\bibinfo
  {volume} {104}},\ \bibinfo {pages} {055036} (\bibinfo {year} {2021})},\
  \Eprint {https://arxiv.org/abs/2102.04474} {arXiv:2102.04474 [hep-ph]}
  \BibitemShut {NoStop}%
\bibitem [{\citenamefont {Arias}\ \emph {et~al.}(2012)\citenamefont {Arias},
  \citenamefont {Cadamuro}, \citenamefont {Goodsell}, \citenamefont {Jaeckel},
  \citenamefont {Redondo},\ and\ \citenamefont {Ringwald}}]{Arias:2012az}%
  \BibitemOpen
  \bibfield  {author} {\bibinfo {author} {\bibfnamefont {P.}~\bibnamefont
  {Arias}}, \bibinfo {author} {\bibfnamefont {D.}~\bibnamefont {Cadamuro}},
  \bibinfo {author} {\bibfnamefont {M.}~\bibnamefont {Goodsell}}, \bibinfo
  {author} {\bibfnamefont {J.}~\bibnamefont {Jaeckel}}, \bibinfo {author}
  {\bibfnamefont {J.}~\bibnamefont {Redondo}},\ and\ \bibinfo {author}
  {\bibfnamefont {A.}~\bibnamefont {Ringwald}},\ }\bibfield  {title} {\bibinfo
  {title} {{WISPy Cold Dark Matter}},\ }\href
  {https://doi.org/10.1088/1475-7516/2012/06/013} {\bibfield  {journal}
  {\bibinfo  {journal} {JCAP}\ }\textbf {\bibinfo {volume} {06}},\ \bibinfo
  {pages} {013}},\ \Eprint {https://arxiv.org/abs/1201.5902} {arXiv:1201.5902
  [hep-ph]} \BibitemShut {NoStop}%
\bibitem [{\citenamefont {Jaeckel}\ \emph {et~al.}(2014)\citenamefont
  {Jaeckel}, \citenamefont {Redondo},\ and\ \citenamefont
  {Ringwald}}]{Jaeckel:2014qea}%
  \BibitemOpen
  \bibfield  {author} {\bibinfo {author} {\bibfnamefont {J.}~\bibnamefont
  {Jaeckel}}, \bibinfo {author} {\bibfnamefont {J.}~\bibnamefont {Redondo}},\
  and\ \bibinfo {author} {\bibfnamefont {A.}~\bibnamefont {Ringwald}},\
  }\bibfield  {title} {\bibinfo {title} {{3.55 keV hint for decaying axionlike
  particle dark matter}},\ }\href {https://doi.org/10.1103/PhysRevD.89.103511}
  {\bibfield  {journal} {\bibinfo  {journal} {Phys. Rev. D}\ }\textbf {\bibinfo
  {volume} {89}},\ \bibinfo {pages} {103511} (\bibinfo {year} {2014})},\
  \Eprint {https://arxiv.org/abs/1402.7335} {arXiv:1402.7335 [hep-ph]}
  \BibitemShut {NoStop}%
\bibitem [{\citenamefont {Ho}\ \emph {et~al.}(2018)\citenamefont {Ho},
  \citenamefont {Saikawa},\ and\ \citenamefont {Takahashi}}]{Ho:2018qur}%
  \BibitemOpen
  \bibfield  {author} {\bibinfo {author} {\bibfnamefont {S.-Y.}\ \bibnamefont
  {Ho}}, \bibinfo {author} {\bibfnamefont {K.}~\bibnamefont {Saikawa}},\ and\
  \bibinfo {author} {\bibfnamefont {F.}~\bibnamefont {Takahashi}},\ }\bibfield
  {title} {\bibinfo {title} {{Enhanced photon coupling of ALP dark matter
  adiabatically converted from the QCD axion}},\ }\href
  {https://doi.org/10.1088/1475-7516/2018/10/042} {\bibfield  {journal}
  {\bibinfo  {journal} {JCAP}\ }\textbf {\bibinfo {volume} {10}},\ \bibinfo
  {pages} {042}},\ \Eprint {https://arxiv.org/abs/1806.09551} {arXiv:1806.09551
  [hep-ph]} \BibitemShut {NoStop}%
\bibitem [{\citenamefont {Hochberg}\ \emph {et~al.}(2018)\citenamefont
  {Hochberg}, \citenamefont {Kuflik}, \citenamefont {Mcgehee}, \citenamefont
  {Murayama},\ and\ \citenamefont {Schutz}}]{Hochberg:2018rjs}%
  \BibitemOpen
  \bibfield  {author} {\bibinfo {author} {\bibfnamefont {Y.}~\bibnamefont
  {Hochberg}}, \bibinfo {author} {\bibfnamefont {E.}~\bibnamefont {Kuflik}},
  \bibinfo {author} {\bibfnamefont {R.}~\bibnamefont {Mcgehee}}, \bibinfo
  {author} {\bibfnamefont {H.}~\bibnamefont {Murayama}},\ and\ \bibinfo
  {author} {\bibfnamefont {K.}~\bibnamefont {Schutz}},\ }\bibfield  {title}
  {\bibinfo {title} {{Strongly interacting massive particles through the axion
  portal}},\ }\href {https://doi.org/10.1103/PhysRevD.98.115031} {\bibfield
  {journal} {\bibinfo  {journal} {Phys. Rev. D}\ }\textbf {\bibinfo {volume}
  {98}},\ \bibinfo {pages} {115031} (\bibinfo {year} {2018})},\ \Eprint
  {https://arxiv.org/abs/1806.10139} {arXiv:1806.10139 [hep-ph]} \BibitemShut
  {NoStop}%
\bibitem [{\citenamefont {Grossman}\ \emph {et~al.}(2002)\citenamefont
  {Grossman}, \citenamefont {Roy},\ and\ \citenamefont
  {Zupan}}]{Grossman:2002by}%
  \BibitemOpen
  \bibfield  {author} {\bibinfo {author} {\bibfnamefont {Y.}~\bibnamefont
  {Grossman}}, \bibinfo {author} {\bibfnamefont {S.}~\bibnamefont {Roy}},\ and\
  \bibinfo {author} {\bibfnamefont {J.}~\bibnamefont {Zupan}},\ }\bibfield
  {title} {\bibinfo {title} {{Effects of initial axion production and photon
  axion oscillation on type Ia supernova dimming}},\ }\href
  {https://doi.org/10.1016/S0370-2693(02)02448-6} {\bibfield  {journal}
  {\bibinfo  {journal} {Phys. Lett. B}\ }\textbf {\bibinfo {volume} {543}},\
  \bibinfo {pages} {23} (\bibinfo {year} {2002})},\ \Eprint
  {https://arxiv.org/abs/hep-ph/0204216} {arXiv:hep-ph/0204216} \BibitemShut
  {NoStop}%
\bibitem [{\citenamefont {Cadamuro}\ and\ \citenamefont
  {Redondo}(2012)}]{Cadamuro:2011fd}%
  \BibitemOpen
  \bibfield  {author} {\bibinfo {author} {\bibfnamefont {D.}~\bibnamefont
  {Cadamuro}}\ and\ \bibinfo {author} {\bibfnamefont {J.}~\bibnamefont
  {Redondo}},\ }\bibfield  {title} {\bibinfo {title} {{Cosmological bounds on
  pseudo Nambu-Goldstone bosons}},\ }\href
  {https://doi.org/10.1088/1475-7516/2012/02/032} {\bibfield  {journal}
  {\bibinfo  {journal} {JCAP}\ }\textbf {\bibinfo {volume} {02}},\ \bibinfo
  {pages} {032}},\ \Eprint {https://arxiv.org/abs/1110.2895} {arXiv:1110.2895
  [hep-ph]} \BibitemShut {NoStop}%
\bibitem [{\citenamefont {Chang}\ \emph {et~al.}(2018)\citenamefont {Chang},
  \citenamefont {Essig},\ and\ \citenamefont {McDermott}}]{Chang:2018rso}%
  \BibitemOpen
  \bibfield  {author} {\bibinfo {author} {\bibfnamefont {J.~H.}\ \bibnamefont
  {Chang}}, \bibinfo {author} {\bibfnamefont {R.}~\bibnamefont {Essig}},\ and\
  \bibinfo {author} {\bibfnamefont {S.~D.}\ \bibnamefont {McDermott}},\
  }\bibfield  {title} {\bibinfo {title} {{Supernova 1987A Constraints on
  Sub-GeV Dark Sectors, Millicharged Particles, the QCD Axion, and an
  Axion-like Particle}},\ }\href {https://doi.org/10.1007/JHEP09(2018)051}
  {\bibfield  {journal} {\bibinfo  {journal} {JHEP}\ }\textbf {\bibinfo
  {volume} {09}},\ \bibinfo {pages} {051}},\ \Eprint
  {https://arxiv.org/abs/1803.00993} {arXiv:1803.00993 [hep-ph]} \BibitemShut
  {NoStop}%
\bibitem [{\citenamefont {Leveille}(1978)}]{LEVEILLE197863}%
  \BibitemOpen
  \bibfield  {author} {\bibinfo {author} {\bibfnamefont {J.~P.}\ \bibnamefont
  {Leveille}},\ }\bibfield  {title} {\bibinfo {title} {The second-order weak
  correction to (g - 2) of the muon in arbitrary gauge models},\ }\href
  {https://doi.org/https://doi.org/10.1016/0550-3213(78)90051-2} {\bibfield
  {journal} {\bibinfo  {journal} {Nuclear Physics B}\ }\textbf {\bibinfo
  {volume} {137}},\ \bibinfo {pages} {63} (\bibinfo {year} {1978})}\BibitemShut
  {NoStop}%
\bibitem [{\citenamefont {Lindner}\ \emph {et~al.}(2018)\citenamefont
  {Lindner}, \citenamefont {Platscher},\ and\ \citenamefont
  {Queiroz}}]{Lindner:2016bgg}%
  \BibitemOpen
  \bibfield  {author} {\bibinfo {author} {\bibfnamefont {M.}~\bibnamefont
  {Lindner}}, \bibinfo {author} {\bibfnamefont {M.}~\bibnamefont {Platscher}},\
  and\ \bibinfo {author} {\bibfnamefont {F.~S.}\ \bibnamefont {Queiroz}},\
  }\bibfield  {title} {\bibinfo {title} {{A Call for New Physics : The Muon
  Anomalous Magnetic Moment and Lepton Flavor Violation}},\ }\href
  {https://doi.org/10.1016/j.physrep.2017.12.001} {\bibfield  {journal}
  {\bibinfo  {journal} {Phys. Rept.}\ }\textbf {\bibinfo {volume} {731}},\
  \bibinfo {pages} {1} (\bibinfo {year} {2018})},\ \Eprint
  {https://arxiv.org/abs/1610.06587} {arXiv:1610.06587 [hep-ph]} \BibitemShut
  {NoStop}%
\bibitem [{\citenamefont {Buttazzo}\ \emph {et~al.}(2021)\citenamefont
  {Buttazzo}, \citenamefont {Panci}, \citenamefont {Teresi},\ and\
  \citenamefont {Ziegler}}]{Buttazzo:2020vfs}%
  \BibitemOpen
  \bibfield  {author} {\bibinfo {author} {\bibfnamefont {D.}~\bibnamefont
  {Buttazzo}}, \bibinfo {author} {\bibfnamefont {P.}~\bibnamefont {Panci}},
  \bibinfo {author} {\bibfnamefont {D.}~\bibnamefont {Teresi}},\ and\ \bibinfo
  {author} {\bibfnamefont {R.}~\bibnamefont {Ziegler}},\ }\bibfield  {title}
  {\bibinfo {title} {{Xenon1T excess from electron recoils of non-relativistic
  Dark Matter}},\ }\href {https://doi.org/10.1016/j.physletb.2021.136310}
  {\bibfield  {journal} {\bibinfo  {journal} {Phys. Lett. B}\ }\textbf
  {\bibinfo {volume} {817}},\ \bibinfo {pages} {136310} (\bibinfo {year}
  {2021})},\ \Eprint {https://arxiv.org/abs/2011.08919} {arXiv:2011.08919
  [hep-ph]} \BibitemShut {NoStop}%
\bibitem [{\citenamefont {Mimasu}\ and\ \citenamefont
  {Sanz}(2015)}]{Mimasu:2014nea}%
  \BibitemOpen
  \bibfield  {author} {\bibinfo {author} {\bibfnamefont {K.}~\bibnamefont
  {Mimasu}}\ and\ \bibinfo {author} {\bibfnamefont {V.}~\bibnamefont {Sanz}},\
  }\bibfield  {title} {\bibinfo {title} {{ALPs at Colliders}},\ }\href
  {https://doi.org/10.1007/JHEP06(2015)173} {\bibfield  {journal} {\bibinfo
  {journal} {JHEP}\ }\textbf {\bibinfo {volume} {06}},\ \bibinfo {pages}
  {173}},\ \Eprint {https://arxiv.org/abs/1409.4792} {arXiv:1409.4792 [hep-ph]}
  \BibitemShut {NoStop}%
\bibitem [{\citenamefont {Jaeckel}\ and\ \citenamefont
  {Spannowsky}(2016)}]{Jaeckel:2015jla}%
  \BibitemOpen
  \bibfield  {author} {\bibinfo {author} {\bibfnamefont {J.}~\bibnamefont
  {Jaeckel}}\ and\ \bibinfo {author} {\bibfnamefont {M.}~\bibnamefont
  {Spannowsky}},\ }\bibfield  {title} {\bibinfo {title} {{Probing MeV to 90 GeV
  axion-like particles with LEP and LHC}},\ }\href
  {https://doi.org/10.1016/j.physletb.2015.12.037} {\bibfield  {journal}
  {\bibinfo  {journal} {Phys. Lett. B}\ }\textbf {\bibinfo {volume} {753}},\
  \bibinfo {pages} {482} (\bibinfo {year} {2016})},\ \Eprint
  {https://arxiv.org/abs/1509.00476} {arXiv:1509.00476 [hep-ph]} \BibitemShut
  {NoStop}%
\bibitem [{\citenamefont {Lees}\ \emph {et~al.}(2016)\citenamefont {Lees} \emph
  {et~al.}}]{BaBar:2016sci}%
  \BibitemOpen
  \bibfield  {author} {\bibinfo {author} {\bibfnamefont {J.~P.}\ \bibnamefont
  {Lees}} \emph {et~al.} (\bibinfo {collaboration} {BaBar}),\ }\bibfield
  {title} {\bibinfo {title} {{Search for a muonic dark force at BABAR}},\
  }\href {https://doi.org/10.1103/PhysRevD.94.011102} {\bibfield  {journal}
  {\bibinfo  {journal} {Phys. Rev. D}\ }\textbf {\bibinfo {volume} {94}},\
  \bibinfo {pages} {011102} (\bibinfo {year} {2016})},\ \Eprint
  {https://arxiv.org/abs/1606.03501} {arXiv:1606.03501 [hep-ex]} \BibitemShut
  {NoStop}%
\bibitem [{\citenamefont {Adachi}\ \emph {et~al.}(2024)\citenamefont {Adachi}
  \emph {et~al.}}]{Belle-II:2024wtd}%
  \BibitemOpen
  \bibfield  {author} {\bibinfo {author} {\bibfnamefont {I.}~\bibnamefont
  {Adachi}} \emph {et~al.} (\bibinfo {collaboration} {Belle-II}),\ }\bibfield
  {title} {\bibinfo {title} {{Search for a \ensuremath{\mu}+\ensuremath{\mu}-
  resonance in four-muon final states at Belle II}},\ }\href
  {https://doi.org/10.1103/PhysRevD.109.112015} {\bibfield  {journal} {\bibinfo
   {journal} {Phys. Rev. D}\ }\textbf {\bibinfo {volume} {109}},\ \bibinfo
  {pages} {112015} (\bibinfo {year} {2024})},\ \Eprint
  {https://arxiv.org/abs/2403.02841} {arXiv:2403.02841 [hep-ex]} \BibitemShut
  {NoStop}%
\bibitem [{\citenamefont {Dolan}\ \emph {et~al.}(2015)\citenamefont {Dolan},
  \citenamefont {Kahlhoefer}, \citenamefont {McCabe},\ and\ \citenamefont
  {Schmidt-Hoberg}}]{Dolan:2014ska}%
  \BibitemOpen
  \bibfield  {author} {\bibinfo {author} {\bibfnamefont {M.~J.}\ \bibnamefont
  {Dolan}}, \bibinfo {author} {\bibfnamefont {F.}~\bibnamefont {Kahlhoefer}},
  \bibinfo {author} {\bibfnamefont {C.}~\bibnamefont {McCabe}},\ and\ \bibinfo
  {author} {\bibfnamefont {K.}~\bibnamefont {Schmidt-Hoberg}},\ }\bibfield
  {title} {\bibinfo {title} {{A taste of dark matter: Flavour constraints on
  pseudoscalar mediators}},\ }\href {https://doi.org/10.1007/JHEP03(2015)171}
  {\bibfield  {journal} {\bibinfo  {journal} {JHEP}\ }\textbf {\bibinfo
  {volume} {03}},\ \bibinfo {pages} {171}},\ \bibinfo {note} {[Erratum: JHEP
  07, 103 (2015)]},\ \Eprint {https://arxiv.org/abs/1412.5174} {arXiv:1412.5174
  [hep-ph]} \BibitemShut {NoStop}%
\bibitem [{\citenamefont {Ametller}\ \emph {et~al.}(1983)\citenamefont
  {Ametller}, \citenamefont {Bergstrom}, \citenamefont {Bramon},\ and\
  \citenamefont {Masso}}]{Ametller:1983ec}%
  \BibitemOpen
  \bibfield  {author} {\bibinfo {author} {\bibfnamefont {L.}~\bibnamefont
  {Ametller}}, \bibinfo {author} {\bibfnamefont {L.}~\bibnamefont {Bergstrom}},
  \bibinfo {author} {\bibfnamefont {A.}~\bibnamefont {Bramon}},\ and\ \bibinfo
  {author} {\bibfnamefont {E.}~\bibnamefont {Masso}},\ }\bibfield  {title}
  {\bibinfo {title} {{The Quark Triangle: Application to Pion and $\eta$
  Decays}},\ }\href {https://doi.org/10.1016/0550-3213(83)90326-7} {\bibfield
  {journal} {\bibinfo  {journal} {Nucl. Phys. B}\ }\textbf {\bibinfo {volume}
  {228}},\ \bibinfo {pages} {301} (\bibinfo {year} {1983})}\BibitemShut
  {NoStop}%
\end{thebibliography}%
\end{document}